\documentclass{jfm}

\usepackage{graphicx}
\usepackage{natbib}

\ifCUPmtlplainloaded \else
  \checkfont{eurm10}
  \iffontfound
    \IfFileExists{upmath.sty}
      {\typeout{^^JFound AMS Euler Roman fonts on the system,
                   using the 'upmath' package.^^J}%
       \usepackage{upmath}}
      {\typeout{^^JFound AMS Euler Roman fonts on the system, but you
                   dont seem to have the}%
       \typeout{'upmath' package installed. JFM.cls can take advantage
                 of these fonts,^^Jif you use 'upmath' package.^^J}%
      }
  \else
  \fi
\fi

\ifCUPmtlplainloaded \else
  \checkfont{msam10}
  \iffontfound
    \IfFileExists{amssymb.sty}
      {\typeout{^^JFound AMS Symbol fonts on the system, using the
                'amssymb' package.^^J}%
       \usepackage{amssymb}%
         \let\leq=\leqslant
         
      }{}
  \fi
\fi

\ifCUPmtlplainloaded \else
  \IfFileExists{amsbsy.sty}
    {\typeout{^^JFound the 'amsbsy' package on the system, using it.^^J}%
     \usepackage{amsbsy}}
    {\providecommand\boldsymbol[1]{\mbox{\boldmath $##1$}}}
\fi

\newsavebox{\astrutbox}
\sbox{\astrutbox}{\rule[-5pt]{0pt}{20pt}}

\title[Oscillatory superfluid Ekman pumping]{Oscillatory superfluid Ekman pumping in Helium II and neutron stars}

\author[C. A. van Eysden]%
{C. Anthony van Eysden$^{1,2}$%
  \thanks{Email address for correspondence: anthonyvaneysden@montana.edu}}

\affiliation{$^1$Department of Physics, Montana State University, Bozeman, Montana, 
59717, USA \\[\affilskip]
$^2$ Nordita, KTH Royal Institute of Technology and Stockholm University, Roslagstullsbacken 23, SE-10691 Stockholm, Sweden}

\pubyear{2010}
\volume{650}
\pagerange{119--126}
\date{?; revised ?; accepted ?. - To be entered by editorial office}
\begin{document}

\maketitle

\begin{abstract}
The linear response of a superfluid, rotating uniformly in a cylindrical container and threaded with a large number of vortex lines, to an impulsive increase in the angular velocity of the container is investigated.  
At zero temperature and with perfect pinning of vortices to the top and bottom of the container, we demonstrate that the system oscillates persistently with a frequency proportional to the vortex line tension parameter to the quarter power.  
This low-frequency mode is generated by a secondary flow analogous to classical Ekman pumping that is periodically reversed by the vortex tension in the boundary layers.  
We compare analytic solutions to the two-fluid equations of \citet{cha86} with the spin-up experiments of \citet{tsa80} in helium II and find the frequency agrees within a factor of four, although the experiment is not perfectly suited to the application of the linear theory.
We argue that this oscillatory Ekman pumping mode, and not Tkachenko modes provide a natural explanation for the observed oscillation.
In neutron stars, the oscillation period depends on the pinning interaction between neutron vortices and flux tubes in the outer core.  Using a simplified pinning model, we demonstrate that strong pinning can accommodate modes with periods of days to years, which are only weakly damped by mutual friction over longer timescales.

\end{abstract}

\begin{keywords}
Authors should not enter keywords on the manuscript, as these must be chosen by the author during the online submission process and will then be added during the typesetting process (see http://journals.cambridge.org/data/\linebreak[3]relatedlink/jfm-\linebreak[3]keywords.pdf for the full list)
\end{keywords}

\section{Introduction}
\label{sec1}

The linear response of a rapidly rotating Navier-Stokes fluid to an impulsive increase in angular velocity of its container, or `spin-up' has been extensively studied.  In the seminal work of \citet{gre63}, three phases were identified: formation of a viscous boundary layer, Ekman pumping, and the damping of residual inertial oscillations by viscosity. Co-rotation between the interior fluid and container is established by Ekman pumping, during which a secondary flow recycles fluid from the boundary layers into the interior.  The time-scale for the spin-up, known as the Ekman time, is proportional to the viscosity to the minus half power, and achieves co-rotation much faster than viscous diffusion.  Since the work of \citet{gre63}, spin-up has been studied in fluids with density stratification \citep{ped67,wal69}, stratified and compressible fluids \citep{abn96,van08a} magnetized plasmas \citep{lop71a,eas79a,van14b}, and multi-component fluids \citep{ung90,amb93} and different geometries \citep{cla71,van13,van14a}.  For a review of spin-up in classical fluids, the reader is referred to \citep{ben74}.

The spin-up of superfluids like helium II has also received considerable attention both experimentally \citep{tsa80} and theoretically \citep{cam82,ada85,per90} with the aim of shedding light on the interaction between the quantized vortices and the container walls.
A rigorous treatment was provided by \citet{rei93}, who solved the two-fluid equations of \citet{cha86} between {\it slowly accelerating} parallel plates, and showed that the Ekman time is reduced by a factor depending on the superfluid mutual friction coefficients and the normal fluid density fraction.  This was generalized to the {\it impulsive} acceleration of containers of arbitrary shape \citep{van13} and to include the self-consistent response of the container \citep{van14a} to facilitate comparison with helium II experiments \citep{van11a,van11b}.  However, theoretical work to date on the spin-up of two-component superfluids such as helium II assumes smooth-walled containers and neglects the effects of pinning.

The spin up of helium II at zero temperature, where the normal fluid component vanishes and cannot facilitate Ekman pumping was also investigated by \citet{rei93}.  By introducing a frictional force to account for the sliding of vortex lines at the boundary, \citet{rei93} showed that the superfluid spins up via an Ekman-like secondary flow with a timescale that depends on the strength of the frictional force.  However, only the response of a single-component fluid to {\it slowly accelerating} parallel plates was studied.

There is good reason to expect that the impulsive acceleration problem produces different physics than that of the slowly accelerating parallel plates.  
When an ideal magnetohydrodynamic plasma is slowly accelerated (i.e., the time-scale for acceleration of the container is the slower than the Alfv\'en crossing time), the plasma moves with the container as a rigid body.  However, in response to an impulsive acceleration, Alfv\'en waves are excited that produce persistent oscillations of the container and the plasma \citep{van14b}.  A final state of co-rotation between the container and plasma is inconsistent with energy conservation if the fluid is dissipation-less, hence the system oscillates persistently.  Similar arguments apply for a superfluid at zero temperature rotating uniformly with a high density of vortex lines.

Two primary motivations exist for this study.  The first is a series of experiments performed by \citet{tsa80}, in which the angular velocity of containers filled with uniformly rotating helium II was impulsively increased. The subsequent motion of the container, responding freely to the hydrodynamic torque of the contained fluid, was then recorded.   In experiments where the container was coated with powder to facilitate vortex pinning, a sinusoidal oscillation of the container was observed.
These oscillations are have been interpreted as Tkachenko modes, however the dependence of the oscillation period on vessel radius predicted for these modes is inconsistent with measurements, whereas the columnar nature of the Ekman pumping mechanism presented here is not.

The second application is the recovery of pulsar glitches; the original motivation for experiments of \citet{tsa80}.  Glitches are tiny, impulsive increases in the rotation frequency of the neutron star crust, typically followed by a quasi-exponential recovery that is believed to be associated with the response of the interior superfluid \cite{bay69b}. Although the superfluid neutron vortices are believed to pin to lattice sites in the crust and to flux tubes arising from type II superconductivity of protons in the outer core, the Ekman pumping mechanism identified here may give rise to long-period oscillations if vortex creep is active.

In this paper, we revisit the problem of superfluid spin-up, by solving for the response of a uniformly rotating superfluid to an {\it impulsive} acceleration of the container.  
We focus on cylindrical geometry, where the vortices are strongly pinned to the top and bottom of the container.
We apply the traditional Laplace transform techniques of \citet{gre63}, but solve self-consistently for the motion of the container as in \citet{van14a} and \citet{van14b}.   
We demonstrate the presence of a low-frequency oscillation mode, with a period that scales as the vortex line parameter to the one-fourth power.
This mode arises in the fast rotation limit, analogous to classical  Ekman pumping.
A secondary flow is present, which is periodically reversed by the tension in the vortex lines as they are sheared in the boundary layer.  By solving the two-fluid equations of \citet{cha86}, we predict an oscillation period of 10 seconds in helium II, compared with that of 40 seconds observed in the experiments of \citet{tsa80}.  A direct comparison cannot be made because the experiment is not perfectly suited for the application of the linear theory of \citet{cha86}.
Using a simplified pinning model, we predict that oscillations in neutron stars can have periods of days to years if the pinning of neutron vortices to flux tubes in the core is strong.  These oscillations are weakly damped and expected to be observable over timescales much longer than the period.

The paper is structured as follows.  In \S\ref{sec2} we linearize the equations of \citet{cha86} using a common Ekman pumping ansatz.  These equations are solved in  \S\ref{sec3} for a superfluid at $T=0$ and perfect vortex pinning of vortices at the top and bottom of a cylindrical container, where it is shown that the solution is oscillatory and has an Ekman-like secondary circulation.  In \S\ref{sec4}, the full two-fluid equations are solved and compared with the experiments of \citet{tsa80} in helium II.  Neutron star applications are considered in \S\ref{sec5}.  In \S\ref{sec6} we summarize our conclusions.  Technical details of the solution are presented in the appendices along with some important limiting cases.

\section{Governing equations}\label{sec2}

A convenient description of superfluids such as helium II, rotating with a high density of vortex lines is given by the Hall-Vinen-Bekharevich-Khalatnikov (HVBK) equations.  
The fluid comprises two components: a `normal' component, denoted by subscript $n$, with viscosity $\eta$, and an inviscid `superfluid' component, denoted by subscript $s$.
\footnote{In a neutron star, the `normal' and `superfluid' components refer to a proton-electron plasma and neutron superfluid, respectively.  This is discussed further in \S\ref{sec5}. }
Under rotation the inviscid component forms a dense array of quantized vortices, which are smooth-averaged in the hydrodynamic approximation, endowing the inviscid component with a macroscopic vorticity.  The vortices mediate interactions between the normal and superfluid components, giving rise to a mutual friction force.   An informative introduction to the HVBK equations is provided in \citet{hen00}.
A more general set of equations, derived by \citet{bay83,cha86}, includes restoring forces experienced by the vortex lattice when vortices are displaced from their equilibrium configuration.  This force produces Tkachenko oscillations, which are expected to be observed in superfluid experiments.
Assuming that the fluid is incompressible, the equations of \citet{cha86}, written in HVBK form and in the laboratory frame are
\begin{eqnarray}
\frac{\partial \boldsymbol{v}_n}{\partial t}+ \boldsymbol{v}_n\cdot \boldsymbol{\nabla}\boldsymbol{v}_n&=&-\boldsymbol{\nabla} p_n + \frac{\rho_s}{\rho} \boldsymbol{F}+\nu_n \nabla^2 \boldsymbol{v}_n\,, \label{eq2.1} \\
\frac{\partial \boldsymbol{v}_s}{\partial t}+ \boldsymbol{v}_s\cdot \boldsymbol{\nabla}\boldsymbol{v}_s&=&-\boldsymbol{\nabla} p_s -\boldsymbol{t}-\boldsymbol{\sigma}-\frac{\rho_n}{\rho} \boldsymbol{F}\,, \label{eq2.2} \\
\boldsymbol{\nabla}\cdot\boldsymbol{v}_n&=&0 \label{eq2.3}\,, \\
\boldsymbol{\nabla}\cdot\boldsymbol{v}_s&=&0 \label{eq2.4}\,,
\end{eqnarray}
where $\boldsymbol{v}_{n,s}$, $\rho_{n,s}$ and $p_{n,s}$ are the macroscopic velocities, densities and reduced pressures of the normal and superfluid components, respectively.
Throughout our analysis, both the normal and superfluid components are considered incompressible; the validity of this assumption is assessed in \S\ref{sec4} and \S\ref{sec5} for helium II and neutron stars respectively.
The mutual friction force is
\begin{eqnarray}
 \boldsymbol{F}&=&\frac{1}{2}B \hat{\omega}_s\times\left[\boldsymbol{\omega}_s\times\left(\boldsymbol{v}_n-\boldsymbol{v}_s\right)-\boldsymbol{t}-\boldsymbol{\sigma} \right] \nonumber \\
 &&+\frac{1}{2}B' \left[\boldsymbol{\omega}_s\times\left(\boldsymbol{v}_n-\boldsymbol{v}_s\right)-\boldsymbol{t}-\boldsymbol{\sigma} \right]\,, \label{eq2.5}
\end{eqnarray}
and
\begin{equation} \label{eq2.6}
 \boldsymbol{\omega}_s=\boldsymbol{\nabla}\times\boldsymbol{v}_s\,,
\end{equation}
is the {\it macroscopic} vorticity of the superfluid and $\hat{\omega}_s=\boldsymbol{\omega}_s/|\boldsymbol{\omega}_s|$
is the vortex line direction vector, and $B$ and $B'$ are dimensionless mutual friction coefficients.  
The vortex tension force per unit mass is
\begin{equation} \label{eq2.7}
 \boldsymbol{t}=\nu_s \boldsymbol{\omega}_s \times \left(\boldsymbol{\nabla} \times \hat{\boldsymbol{\omega}}_s \right)\,.
\end{equation}
where the vortex line tension parameter is given by
\begin{equation}
  \nu_s= \frac{\Gamma}{4 \pi}  \log\left(\frac{b_0}{a_0}\right) \,. \label{eq2.8} 
\end{equation}
Here $\Gamma=\hbar \pi/m$ is the quantum of circulation, $m$ is the mass of one helium atom, $b_0$ is the inter-vortex spacing and $a_0$ is the size of the vortex core.
The kinematic viscosity $\nu_n$ is defined as the shear viscosity divided by the {\it normal fluid} density, $\eta/\rho_n$.
The final parameter $\boldsymbol{\sigma}$ in (\ref{eq2.1})--(\ref{eq2.4}) comes from the theory of \citet{bay83} and describes the restoring force of the vortex lattice in response to shear deformations. It has the form
\begin{equation} \label{eq2.9} 
 \boldsymbol{\sigma}=\frac{\hbar |\boldsymbol{\omega}_s|}{8 m}\left[2\boldsymbol{\nabla}_{\perp}\left(\boldsymbol{\nabla}\cdot\boldsymbol{\xi}\right)-\boldsymbol{\nabla}_{\perp}^2 \boldsymbol{\xi}\right]\,,
\end{equation}
where $\boldsymbol{\xi}$ is the vortex line displacement vector  and $\boldsymbol{\nabla}_{\perp}$ is the two-dimensional gradient operator.  Both $\boldsymbol{\xi}$ and $\boldsymbol{\nabla}_{\perp}$ are two-dimensional in the plane orthogonal to the angular velocity of the background superfluid flow, i.e., $\hat{\omega}_s\cdot\boldsymbol{\xi}=\hat{\omega}_s\cdot\boldsymbol{\nabla}_{\perp}=0$.
Equation (\ref{eq2.9}) only applies to linear deformations of a rectilinear vortex array, hence (\ref{eq2.1})--(\ref{eq2.4}) are only valid in the linear approximation when $\boldsymbol{\sigma}$ is included.
When $\boldsymbol{\sigma}=0$, (\ref{eq2.1})--(\ref{eq2.4}) are the HVBK equations, which describe non-linear flows including quasi-classical turbulence \citep{hen95,per08}.

The vortex lines obey the vorticity conservation law
\begin{equation}
  \frac{\partial \boldsymbol{ \omega}_s}{\partial t}=\nabla\times\left(\boldsymbol{v}_L \times \boldsymbol{ \omega}_s  \right) \,, \label{eq2.10}
\end{equation}
where in the linear approximation the perturbation to the vortex line velocity is given by $\partial \xi/\partial t$.  
Taking the curl of (\ref{eq2.2}) and comparing the result with (\ref{eq2.10}) gives the equation of motion for the vortex lines
\begin{equation}
   \boldsymbol{ \omega}_s\times\left(\boldsymbol{v}_L-\boldsymbol{ v}_s \right)=\boldsymbol{t}+\boldsymbol{\sigma}+\frac{\rho_n}{\rho} \boldsymbol{ F}\,. \label{eq2.12}
\end{equation}

Equations (\ref{eq2.1}) and (\ref{eq2.2}) can be combined into an equation for the total fluid, 
\begin{eqnarray}
\frac{\partial }{\partial t} \left( \rho_n\boldsymbol{v}_n+\rho_s \boldsymbol{v}_s \right)+ \nabla_j\left( \rho_n v_{ni} v_{nj} +\rho_s v_{s i}v_{s j} \right)&=&\nabla_j T_{ij}\,, \label{eq2.13}
\end{eqnarray}
where
\begin{equation}
 T_{ij}=-p \delta_{ij}+T^v_{ij}+T^s_{ij} +T^t_{ij}\,,\label{eq2.14} 
\end{equation}
The contributions to the stress are the total pressure
\begin{equation} \label{eq2.18} 
  p=\rho_n p_n+\rho_s p_s- \rho_s \nu_s |\boldsymbol{\omega}_s| \,.
\end{equation}
the viscous stress
\begin{eqnarray}
 T^v_{ij}&=&\rho_n \nu_n \left(\nabla_i v_j +\nabla_j  v_i\right) \,,\label{eq2.15}
\end{eqnarray}
the vortex line tension,
\begin{eqnarray}
T^s_{ij}&=& \rho_s \nu_s |\boldsymbol{\omega}_s| \left( \hat{\omega}_{s i} \hat{\omega}_{s j} - \delta_{ij}\right) \,,\label{eq2.16}
\end{eqnarray}
and the stress arising from the displacement of vortices from the equilibrium configuration in the lattice,
\begin{eqnarray}
T^t_{ij}&=& \frac{\rho_s \hbar |\boldsymbol{\omega}_s|}{8 m}\left[\nabla_{\perp i} \xi_j+\nabla_{\perp j} \xi_i-3\delta_{ij}\left(\partial_k \xi_k\right)\right]\,. \label{eq2.17}
\end{eqnarray}
The term (\ref{eq2.17}) is responsible for Tkachenko oscillations.

To study the coupled response of a superfluid and
its container, we consider two infinite parallel plates with separation $2 L$. 
This geometry has a long history in the
study of the spin up of rapidly rotating fluids in geophysics \citep{gre63, ped67,wal69}, magnetized plasmas \citep{lop71a,eas79b,van14b} , condensed matter \citep{rei93} and astrophysics \citep{abn96}. 
At times $t < 0$, the superfluid and its container rotate rigidly and uniformly about the cylindrical axis with angular velocity $\Omega$. 
At time $t=0$, the magnitude of the angular velocity of the container is impulsively increased to $\Omega(1+\epsilon)$, where the $\epsilon \ll 1$ is the Rossby number.
For $t>0$ the container and fluid are left to evolve freely, and we solve self-consistently for the coupled motion of the container and plasma as in previous studies \citep{van13,van14a,van14b}.
The side walls of the container and neglected, however they typically play a secondary role in spin up problems like that considered here.  Side wall effects in helium II experiments are discussed in \S\ref{sec3.3} 

To solve for the motion of the container, we invoke angular momentum conservation between the container and fluid\begin{equation}
  I_c \frac{d \boldsymbol{\Omega}_c}{ dt}=-\oint \boldsymbol{x}\times \left(\hat{\boldsymbol{n}} \cdot \mathsfbi{T} \right) {\rm dS}+\boldsymbol{\tau }_{ext} \,, \label{eq2.19}
\end{equation}
where $I_c$ is the moment of inertia of the container, $\hat{\boldsymbol{n}}$ is the unit vector normal to the boundary and ${\rm dS}$ is an element of area on the boundary.  The first term on the right hand side of (\ref{eq2.19}) is the hydrodynamic  torque exerted on the crust by the fluid arising from viscous stresses and stress exerted by the vortex array, where $\mathsfbi{T} $ is given by (\ref{eq2.13}). 
The second term is an external torque which may arise from, e.g., friction in the apparatus for superfluid experiments or the magnetic dipole torque in pulsars.

The normal fluid co-rotates with the container, giving the boundary condition
\begin{eqnarray}
  \boldsymbol{v}_n=\boldsymbol{ \Omega}_c \times \boldsymbol{ x} \,, \label{eq2.20}
\end{eqnarray}
where $\boldsymbol{x}$ is the radial vector and $\boldsymbol{ \Omega_c}(t)$ is the angular velocity of the container, which is a function of time.  
Equation (\ref{eq2.20}) embodies the usual no-slip boundary conditions for viscous flows.
Following \cite{rei93}, for the superfluid we choose the following boundary conditions
\begin{equation}
  \hat{\boldsymbol{n}} \times \left[ \rho_s |\boldsymbol{\omega}_s| L \gamma \left(\boldsymbol{v}_L-\boldsymbol{\Omega}_c\times\boldsymbol{x} \right)\pm \left(\hat{\boldsymbol{n}} \cdot \mathsfbi{T} \right)\right]=0\,, \hspace{10mm} \hat{\boldsymbol{n}} \cdot \boldsymbol{v}_s=0\,, \label{eq2.21}
\end{equation}
on surfaces that intersect vortex lines, where $\hat{\boldsymbol{n}}$ is the unit vector normal to the surface. 
The dimensionless constant $\gamma$ governs the rate of vortex creep at the boundary.  
In the limit $\gamma\rightarrow \infty$, the vortices are pinned to the boundary and when
$\gamma=0$, the vortices exert no stress on the boundary, i.e., they are freely sliding.
For the normal component of the superfluid we require no-penetration.

In cylindrical coordinates $(r,\phi,z)$, the initial conditions are
\begin{eqnarray} 
  \boldsymbol{ v}_{n,s}(0)=r \Omega \hat{\phi}\,, \hspace{1cm} \boldsymbol{ \Omega}_c(0)=\Omega(1+\epsilon)\hat{z} \,, \label{eq2.22}
\end{eqnarray}
i.e., we assume the two fluids are initially co-rotating. Strictly speaking, the initial velocity for the fluid components obey (\ref{eq2.22}) everywhere except in an infinitely thin region adjacent to the boundary where it is spun up by the container.

For $\epsilon \ll 1$, equations (\ref{eq2.1})--(\ref{eq2.22}) can be linearized by perturbing around an equilibrium rotating with uniform angular velocity $\Omega$ about the $z$-axis.  
The external torque is also taken to be aligned with the rotation axis, i.e., $\boldsymbol{ \tau}_{ext}=\tau_{ext} \hat{z}$.
The geometry and initial conditions are axisymmetric, and the resulting flow axisymmetric.
The following substitutions are made for the velocity and pressure fields
\begin{eqnarray}
  \boldsymbol{ v}_{n,s} (r,z,t)&\rightarrow& r \Omega \hat{\phi} +  \epsilon \Omega L \left[ r^*\frac{ \partial \chi_{n,s}}{\partial z^*}\hat{r}+r^*V_{n,s}\hat{\phi}-2   \chi_{n,s} \hat{z}\right]  \,,  \nonumber  \\ 
\boldsymbol{ v}_L(r,z,t)&\rightarrow& r \Omega \hat{\phi} +  \epsilon  \Omega L r^* \left[ \frac{\partial U_\xi}{\partial t} \hat{r}+\frac{\partial V_\xi}{\partial t} \hat{\phi} \right]  \,,  \nonumber  \\ 
p_{n,s}(r,z,t)&\rightarrow &  \frac{ \left( r \Omega \right) ^2}{2}  + \epsilon  \Omega^2 L^2 \left(\frac{ r^{*2} P_{n,s}}{2} +2 Q_{n,s}\right)  \,, \label{eq2.23}
\end{eqnarray}
where $\chi_{n,s}(z^*,t^*)$, $V_{n,s}(z^*,t^*)$, $Q_{n,s}(z^*,t^*)$ and $P_{n,s}(t^*)$ are all dimensionless quantities.
The functions $r \chi_{n,s}$ are stream-functions for the secondary flow.
The asterisked quantities are defined as $r^*=r/L$, $z^*=z/ L$, $t^*=\Omega t$. 
Equation (\ref{eq2.23}) is essentially the ``von-Karman similarity'' form and is typically used in employed in studies of spin-up between parallel plates \citep{gre63,eas79a,rei93,van13}. 
The ansatz (\ref{eq2.23}) automatically satisfies the continuity equations (\ref{eq2.3}) and (\ref{eq2.4}) and the conditions for rotational equilibrium for the background flow.
The radial dependence of the azimuthal velocity is motivated by the boundary conditions (\ref{eq2.20}), and chosen for the other variables such that $r$ vanishes from the resulting equations in the most general way.
Under these assumptions, the only non-vanishing component of the external torque is in the $\hat{z}$ direction, hence
\begin{eqnarray}
 \boldsymbol{ \Omega}_c(t^*)&\rightarrow & \Omega \hat{z} + \epsilon \Omega f(t^*) \hat{z} \label {eq2.24} \,,
\end{eqnarray}
where $f$ is a function of $t^*$ only.
Henceforth, we drop the asterisk notation so that all variables are dimensionless.

Substituting (\ref{eq2.23}) into the normal fluid momentum equation (\ref{eq2.1}), we obtain
\begin{eqnarray}
0&=&\left( \frac{\partial }{\partial t}  -  E \frac{\partial^2}{\partial z^2} +\frac{\rho_s B}{\rho} \right) \frac{\partial \chi_n}{\partial z}  - \left(2-\frac{\rho_s B'}{\rho} \right) V_n \nonumber \\
&& -\frac{\rho_s B}{2\rho}\left(2 -E_s\frac{\partial^2}{\partial z^2}\right) \frac{\partial \chi_s}{\partial z}-\frac{\rho_s B'}{2\rho}\left(2-E_s\frac{\partial^2}{\partial z^2}\right)V_s + P_n \, , \label{eq2.25} \\
0&=&\left( \frac{\partial }{\partial t} - E \frac{\partial^2}{\partial z^2}+ \frac{\rho_s B}{\rho} \right)V_n + \left(2 -\frac{\rho_s B'}{\rho} \right)\frac{\partial \chi_n}{\partial z} \nonumber \\
&& -\frac{\rho_s B}{ 2 \rho} \left( 2 - E_s \frac{\partial^2}{\partial z^2}\right)V_s + \frac{\rho_s B'}{2 \rho} \left( 2 - E_s \frac{\partial^2}{\partial z^2}\right)\frac{\partial \chi_s}{\partial z}\, ,  \label{eq2.26} \\
0&=& \left( \frac{\partial }{\partial t}  -  E \frac{\partial^2}{\partial z^2} \right) \chi_n  -\frac{\partial Q_n}{\partial z} \, , \label{eq2.27}
\end{eqnarray}
and from (\ref{eq2.2}) we have for the superfluid
\begin{eqnarray}
0&=&\left[ \frac{\partial }{\partial t}  +\frac{\rho_n B}{2 \rho} \left(2 - E_s \frac{\partial^2}{\partial z^2}  \right)\right] \frac{\partial \chi_s}{\partial z} - \left(1-\frac{\rho_n B' }{2\rho}\right) \left(2 - E_s \frac{\partial^2 }{\partial z^2}\right) V_s \nonumber \\
&&- \frac{\rho_n B}{\rho} \frac{\partial \chi_n}{\partial z}  -\frac{\rho_n B' }{\rho} V_n +P_s \, , \label{eq2.28}\\
0&=&\left[ \frac{\partial }{\partial t} +\frac{\rho_n B}{2\rho}\left(2-E_s \frac{\partial^2 }{\partial z^2} \right) \right] V_s +\left(1 -\frac{\rho_n B'}{2\rho} \right) \left(2-E_s \frac{\partial^2}{\partial z^2}\right)  \frac{\partial \chi_s}{\partial z} \nonumber \\
&& -\frac{\rho_n B}{\rho} V_n + \frac{\rho_n B'}{\rho}  \frac{\partial \chi_n}{\partial z}\, , \label{eq2.29} \\
  0&=& \frac{\partial \chi_s}{\partial t} - \frac{\partial Q_s}{\partial z}  \, , \label{eq2.30}
\end{eqnarray}
where we define the dimensionless parameters
\begin{eqnarray}
E=\frac{\nu_n}{ L^2 \Omega} \,,\hspace{1cm} E_s=\frac{\nu_s}{L^2 \Omega}\,. \label{eq2.31}
\end{eqnarray}
The dimensionless parameter $E$ is the Ekman number and is a ratio of viscous forces and the rotational inertia in the flow.  
By analogy, we define the superfluid Ekman number $E_s$, which is a ratio of the vortex line tension and rotational inertia in the flow.   
Equation (\ref{eq2.19}) becomes
\begin{eqnarray}
 \frac{d f}{dt}&=&\mp \frac{K}{\rho}  \left(\rho_s E_s \frac{\partial^2 \chi_s}{\partial z^2}+\rho_n E \frac{\partial V_n}{\partial z} \right) \nonumber \\
 &&+\alpha \left(1+K\right)\,, \label{eq2.32}
\end{eqnarray}
at $z=\pm 1$, where we define
\begin{equation}
  K=\frac{ \pi \rho R^4 L}{I_c}\,, \hspace{1cm}  \alpha=\frac{\tau_{ext}}{\epsilon \Omega^2 I_{tot}}\,. \label{eq2.33}
\end{equation}
In (\ref{eq2.32}), the dimensionless parameter $K$ denotes the ratio of the moments of inertia of the fluid and container, and $\alpha$ is the dimensionless external torque where $I_{tot}=I_c (1+K)$ is the total moment of inertia of the fluid and container.

The boundary conditions for the normal fluid (\ref{eq2.20}) become
\begin{eqnarray}
 V_n- r f  &=&0\, , \label{eq2.34} \\
 \frac{\partial \chi_n}{\partial z} &=&0 \, , \label{eq2.35} \\
  \chi_n &=&0\,, \label{eq2.36}
\end{eqnarray}
at $z=\pm1$.
For the superfluid, using the linearized forms of (\ref{eq2.12}) and (\ref{eq2.2}) to eliminate the vortex line velocity and mutual friction, (\ref{eq2.21}) becomes  
\begin{eqnarray}
 \frac{\partial V_s}{\partial t}\pm \frac{E_s}{\gamma} \frac{\partial V_s}{\partial z}  &=&0\, , \label{eq2.37} \\
\frac{\partial }{\partial t}  \frac{\partial \chi_s}{\partial z}-2 f \pm \frac{E_s}{\gamma}\frac{\partial^2 \chi_s}{\partial z^2}+ P_s &=&0 \, , \label{eq2.38} \\
\chi_s &=&0\,, \label{eq2.39}
\end{eqnarray}
at $z=\pm 1$.
The initial conditions (\ref{eq2.22}) become
\begin{eqnarray}
  V_n(z,0)&=&0\,, \\
  V_s(z,0)&=&0\,, \\
  f(0)&=&1\,.
\end{eqnarray}
The vortex line displacements, $\boldsymbol{\xi}$, can be calculated by substituting (\ref{eq2.23}) into the vortex line equation of motion (\ref{eq2.12}) and using (\ref{eq2.2}).  
We find
\begin{eqnarray}
 2\frac{\partial V_\xi}{\partial t}&=& \frac{\partial }{\partial t}\frac{\partial \chi_s }{\partial z}+P_s \,, \\
2\frac{\partial U_\xi}{\partial t}&=& - \frac{\partial V_s }{\partial t} \,.
\end{eqnarray}
For the initial conditions we require
\begin{eqnarray}
  U_\xi(z,0)&=&0\,, \\
  V_\xi(z,0)&=&0\,.
\end{eqnarray}

\section{Pure superfluid}\label{sec3}

\subsection{Exact solution} \label{sec3.1}

To illustrate the Ekman pumping mechanism which is the principal result of this paper, we consider a superfluid at $T=0$, where $\rho_n=0$.
To solve the initial value problem, we take the Laplace transform,
\begin{equation}
  \tilde{X}(z,s)=\int_0^\infty X(z,t) e^{-s t} {\rm d} t\,. \label{eq3.1}
\end{equation}
Equations (\ref{eq2.28}) and (\ref{eq2.29}) are
\begin{eqnarray}
 s \frac{\partial \tilde{\chi}_s}{\partial z} - \left(2 - E_s \frac{\partial^2 }{\partial z^2}\right) \tilde{V}_s  +\tilde{P}_s  &=& 0\, , \label{eq3.2}\\
s \tilde{V}_s + \left(2-E_s \frac{\partial^2}{\partial z^2}\right)  \frac{\partial \tilde{\chi}_s}{\partial z}&=& 0\,  . \label{eq3.3}
\end{eqnarray}
Equation (\ref{eq2.30}) is not required to solve the system, and is only necessary if one desires to calculate $Q_s$ for completeness.  
Taking the derivative of (\ref{eq3.2}) we obtain
\begin{eqnarray}
 s \frac{\partial^2 \tilde{\chi}_s}{\partial z^2} - \left(2 - E_s \frac{\partial^2 }{\partial z^2}\right) \frac{\partial \tilde{V}_s}{\partial z}  &=& 0\, . \label{eq3.4}
\end{eqnarray}
Equations (\ref{eq3.3}) and (\ref{eq3.4}) can be solved for $\tilde{\chi}_s$ and $\tilde{V}_s$;  $\tilde{P}_s$ is then obtained from (\ref{eq3.2}).
The equation for the container, (\ref{eq2.32}) becomes
\begin{eqnarray}
 s \tilde{f}-1&=&\mp K E_s  \left( \frac{\partial^2 \tilde{\chi}_s}{\partial z^2} \right) \,, \label{eq3.5}
\end{eqnarray}
at $z=\pm 1$, and the boundary conditions are
\begin{eqnarray}
 s \tilde{V}_s \pm \frac{E_s}{\gamma} \frac{\partial \tilde{V}_s}{\partial z}  &=&0\, , \label{eq3.6} \\
s \frac{\partial \tilde{\chi}_s}{\partial z}-2 \tilde{f} \pm  \frac{E_s}{\gamma} \frac{\partial^2 \tilde{\chi}_s}{\partial z^2}+ \tilde{P}_s &=&0 \, , \label{eq3.7} \\
\tilde{\chi}_s &=&0\,,\label{eq3.8}
\end{eqnarray}
at $z=\pm 1$.
The solution to (\ref{eq3.3})--(\ref{eq3.8}) is
\begin{eqnarray}
 \tilde{V}_s&=& \frac{2 i \tilde{f}}{\Delta s^2} \left\{ C_+ \left[\kappa_- s \left(\cosh \kappa_- -  \cosh \kappa_- z \right) + (E_s/\gamma) \kappa_-^2 \sinh \kappa_- \right] \right. \,, \nonumber \\
&-& \left. C_- \left[\kappa_+ s \left( \cosh \kappa_+ - \cosh \kappa_+ z \right) + (E_s/\gamma) \kappa_+^2 \sinh \kappa_+ \right] \right\} \,, \label{eq3.9}\\
\tilde{\chi}_s&=&\frac{2 \tilde{f} }{\Delta s} \left[C_+ \left(\sinh \kappa_- z- z \sinh \kappa_-\right) + 
   C_- \left( \sinh \kappa_+ z-  z \sinh \kappa_+ \right)   \right]\,, \label{eq3.10} \\
\tilde{P}_s&=&\frac{2 \tilde{f} }{\Delta s^2} \left(4 + s^2 \right) \left(C_+ \sinh \kappa_- + C_- \sinh \kappa_+ \right) \,,\label{eq3.11}\\
\tilde{f}&=&\frac{\Delta}{\bar{\Delta} s} \label{eq3.12}\,,
\end{eqnarray}
where
\begin{eqnarray}
\kappa_\pm &=& \sqrt{\frac{2 \pm i s}{E_s}}\,,\label{eq3.13} \\
C_\pm &=& \kappa_\pm \cosh \kappa_\pm  + \frac{\left(E_s \kappa_\pm^2 \mp2 i \gamma\right) \sinh \kappa_\pm }{\gamma s} \,, \label{eq3.14}\\
\Delta &=& C_+ \left[ C_- + \frac{2 E_s \kappa_-^2  \sinh \kappa_-}{s^2}\right]+ C_-\left[C_+ +\frac{2 E_s \kappa_+^2  \sinh \kappa_+}{s^2} \right]\,, \label{eq3.15} \\
\bar{\Delta} &=& C_+ \left[ C_- + \frac{2 E_s \kappa_-^2 \left(1+K\right) \sinh \kappa_-}{s^2}\right] \nonumber \\
&+& C_-\left[C_+ +\frac{2 E_s \kappa_+^2 \left(1+K\right) \sinh \kappa_+}{s^2} \right]\,. \label{eq3.16}
\end{eqnarray}
To obtain the response of the container, we must find the inverse Laplace transform of $\tilde{f}$. This is readily done by realizing that there are simple poles at $s=0$ and at the zeroes of $\bar{\Delta}$.  
The result is
\begin{equation}
f(t)=\left[\frac{\Delta}{\bar{\Delta}}\right]_{s=0}+\sum_n R(s_n) e^{s_n t}\,, \label{eq3.17}
\end{equation}
where $s_n$ are the roots of $\bar{\Delta}$ and
\begin{equation}
  R(s) = \frac{\Delta (s)}{s \frac{d \bar{\Delta}}{d s}}\,. \label{eq3.18}
\end{equation}
The sum in (\ref{eq3.17}) implies summing over all the zeroes of $\bar{\Delta}$.
\footnote{The solution to the initial value problem and could also be obtained from a mode expansion and appropriate orthogonality relations, however, the Laplace transform more readily gives the result.}
In general, the eigenvalues occur in conjugate pairs.  For $\gamma \rightarrow \infty$ we have ${\rm Re}(s_n)=0$ and ${\rm Im}[R(s_n)]=0$, and the modes in (\ref{eq3.17}) are purely oscillatory.  When $\gamma$ is finite the friction force results in dissipation and $s_n$ and $R(s_n)$ have real and imaginary components.

Before proceeding, let us consider the relevant range for $E_s$ in hydrodynamic approximation of the HVBK equations.
The vorticity of the superfluid is given by the circulation per vortex multiplied by the number of vortices per unit area, hence
\begin{equation}
  2\Omega= \Gamma \times \frac{N}{\pi R^2}\, \label{eq3.19}
\end{equation}
From (\ref{eq2.8}), (\ref{eq2.31}) and (\ref{eq3.19}) we then obtain
\begin{equation}
 E_s=\frac{R^2}{2 L^2 N}\ln\left(\frac{b_0}{a_0}\right)\,. \label{eq3.20}
\end{equation}
The $\ln$ term is of order unity or an order of magnitude more, so that for a vessel of aspect ratio close to unity we have $E_s\sim 1/N$.  We require $N\gg 1$ for the hydrodynamic approximation to be valid, hence we must have $E_s \ll 1$.

\begin{figure}
  \includegraphics[width=0.9\textwidth]{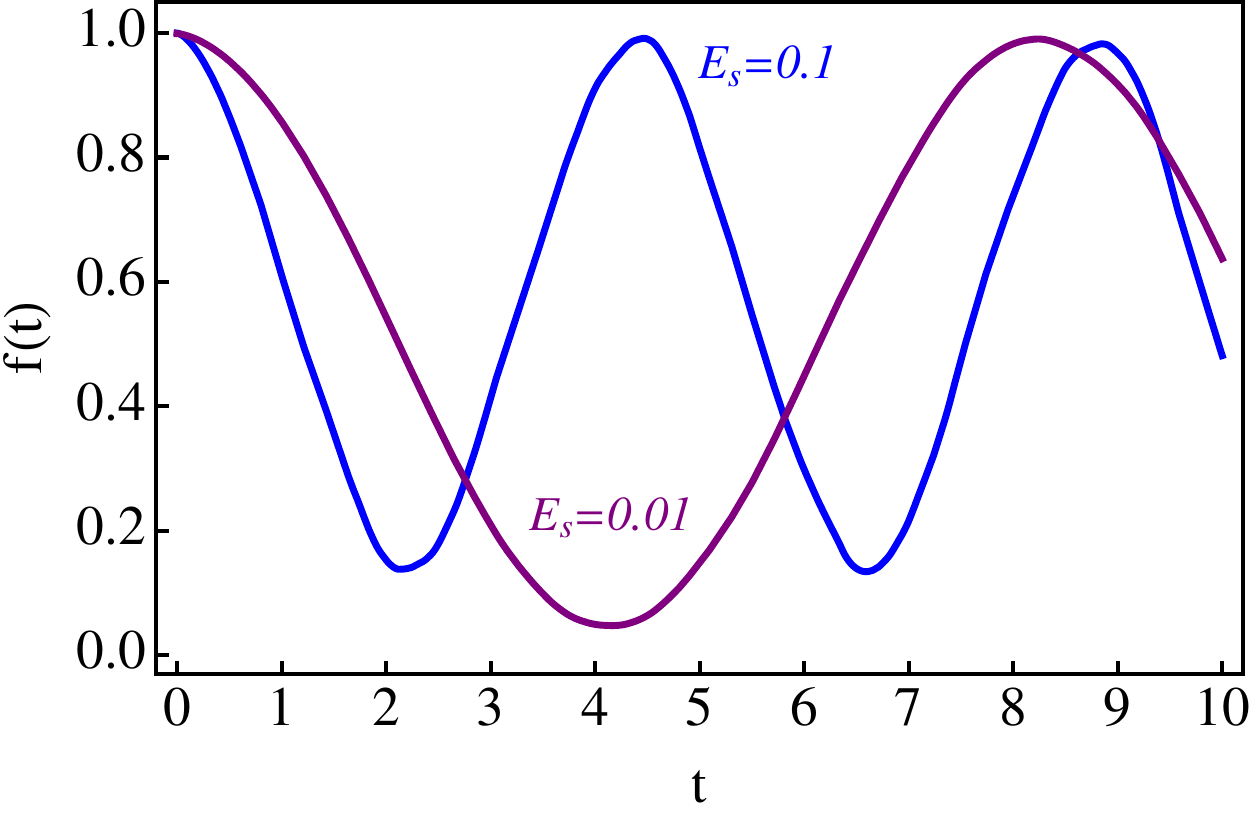}
\caption{Dimensionless angular velocity of the container, $f(t)$, for $E_s=0.1$ (blue curve) and $E_s=0.01$ (purple curve).  We take $K=1$ and $\gamma \rightarrow\infty$.}
\label{fig1}   
\end{figure}

\begin{figure*}
  \includegraphics[width=0.5\textwidth]{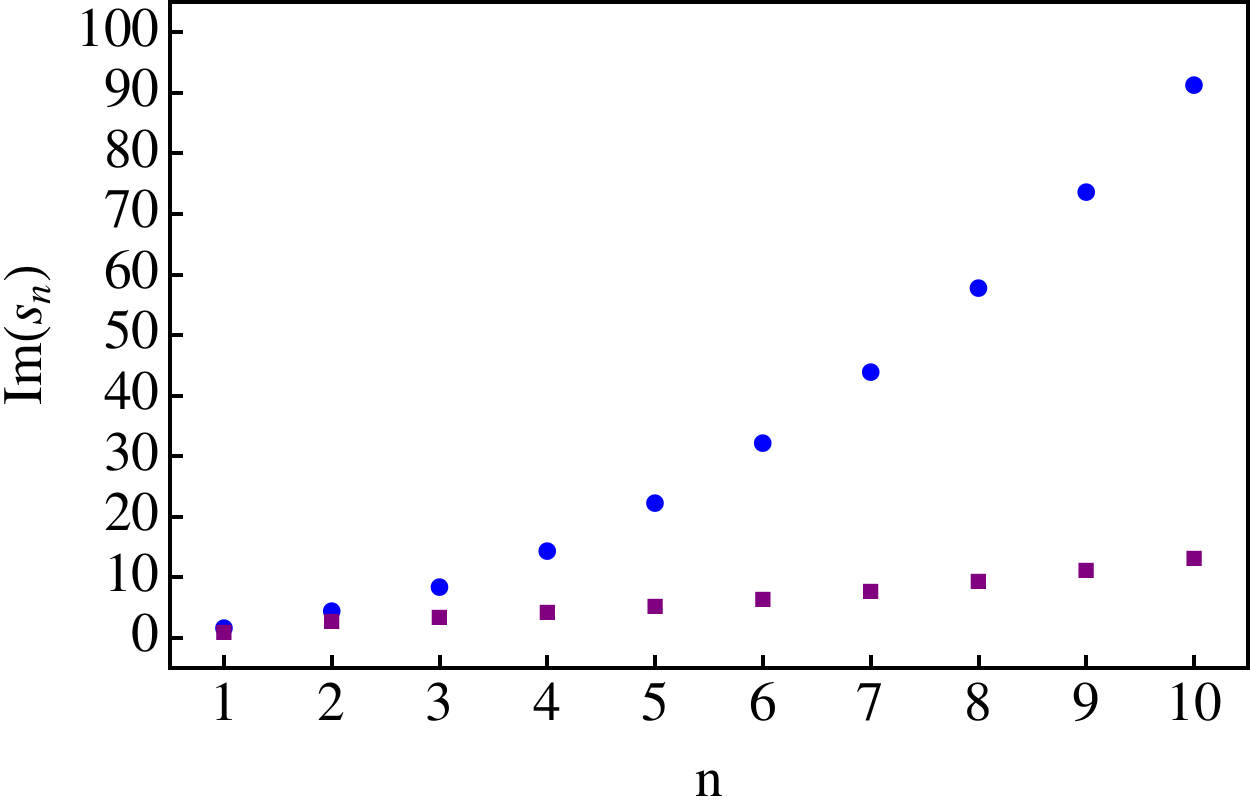} 
  \includegraphics[width=0.5\textwidth]{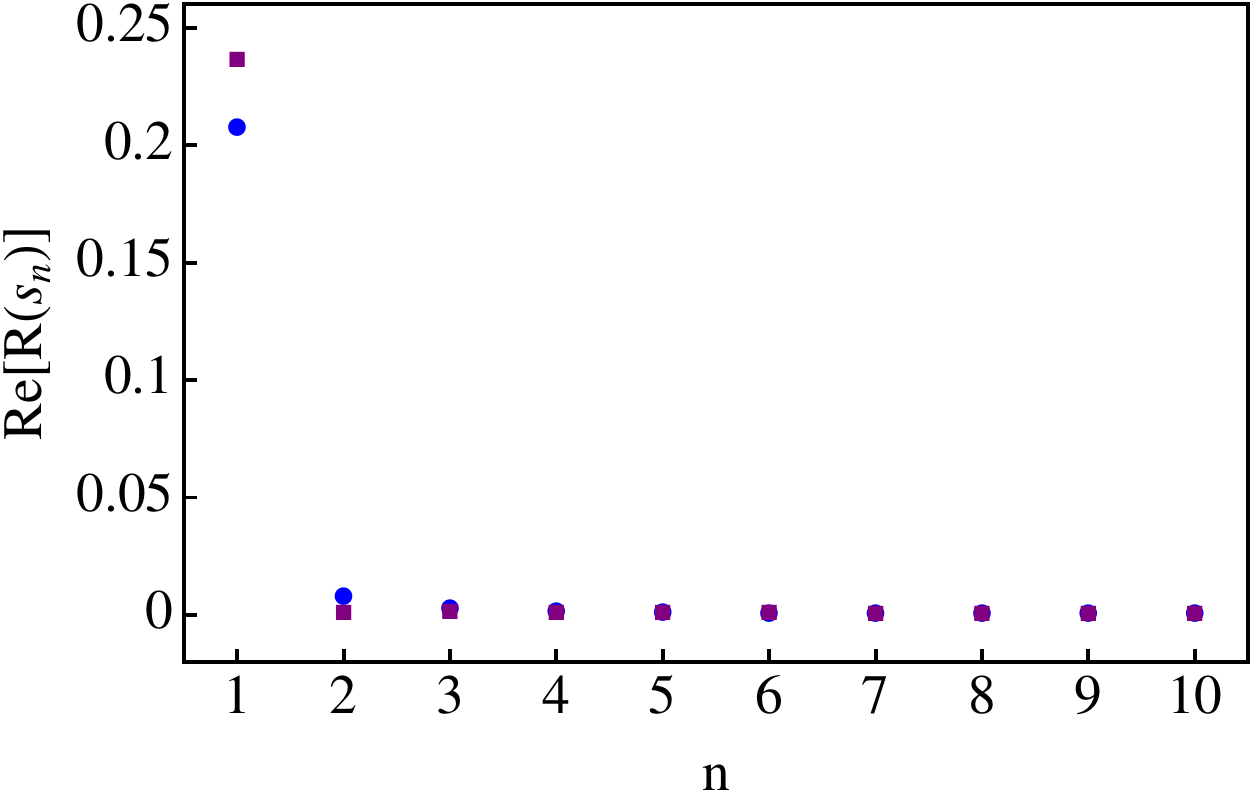} 
\caption{Eigenfrequencies (left panel) and amplitudes (right panel) of the modes present in Figure \ref{fig1}.  Blue circles correspond to $E_s=0.1$, purple squares correspond to $E_s=0.01$.}
\label{fig2}    
\end{figure*}

In Figure \ref{fig1} we plot the response of the container for $E_s=0.1$ (blue curve) and $E_s=0.01$ (purple curve) for $\gamma\rightarrow \infty$ (perfect pinning) and $K=1$ in both cases.  The first 500 eigenmodes of (\ref{eq3.17}) have been plotted.  We find that the container oscillates about the centre of mass of the system in a smooth sinusoidal manner, with a period much greater than the rotation period.  This smooth oscillation can be understood by looking at the modal decomposition of the solution, plotted in Figure \ref{fig2}.  On the left hand side we present the eigenvalues, in increasing order of ${\rm Im}(s_n)$.  On the right hand side we plot the corresponding amplitudes of each mode.  
We see from the right panel that the dominant contribution is from the fundamental mode, with very little contribution from the higher modes.  As $E_s$ is reduces, we find that all the power becomes concentrated in the fundamental mode.
Therefore, for $E_s \ll 1$, the impulsive acceleration of the container only excites the fundamental mode of the system.

To obtain an analytic result for the fundamental mode, we investigate the limit $E_s \ll 1$.  
The frequency of oscillation is much smaller than the rotation frequency, hence we look for solutions in the limit $s\ll 1$.
We then find
\begin{equation}
  \bar{\Delta}\approx \frac{e^{2 k}}{2 s^3 \gamma^2}\left( 2+k \gamma s\right)\left[4 \left(1+K\right)\gamma+2 s+k \gamma s^2\right]\,, \label{eq3.21}
\end{equation}
where $k=\sqrt{2/E_s}\gg 1$.
Equation (\ref{eq3.21}) has zeroes at 
\begin{equation}
  s_\pm=-\frac{1}{k\gamma}\left[ 1\pm \sqrt{1-4\left(1+K\right) k \gamma^2} \right]\,. \label{eq3.22}
\end{equation}
The zero at $s=-2/(k\gamma)$ is also a zero of $\Delta$, and therefore not a pole of $\tilde{f}$.   Evaluating $R(s_\pm)$ we find
\begin{equation}
 f(t)=\frac{1}{1+K}\left[1+\left(\frac{K}{s_+-s_-}\right)\left(s_+ e^{s_-t}-s_- e^{s_+ t}\right)\right]\,. \label{eq3.23}
\end{equation}

We now examine the behavior of (\ref{eq3.23}) by considering two limiting cases.
To recover the oscillatory solution observed in Figure \ref{fig1}, we consider the strong pinning limit $\gamma \gg \left[4\left(1+K\right)k\right]^{-1/2}$.  
From (\ref{eq3.22}) we have
\begin{equation}
  s_\pm=\pm 2 i \sqrt{\frac{1+K}{k}}\,, \label{eq3.24}
\end{equation}
and (\ref{eq3.23}) becomes
\begin{equation}
 f(t)=\frac{1}{1+K}\left[1+K \cos\left(2 \sqrt{\frac{1+K}{k}} t\right)\right]\,. \label{eq3.25}
\end{equation}
We therefore obtain the period of oscillation
\begin{equation} \label{eq3.26}
  t_P=\frac{2\pi }{\Omega | {\rm Im} (s_\pm)| }=\frac{\pi}{\Omega} \sqrt{\frac{k}{1+K}}\,.
\end{equation}
This is the principal result of this paper.  
A single, long time-scale mode has emerged for $E_s \ll 1$, which has a period proportional to $E_s^{-1/4}$.
This is analogous to classical Ekman pumping, where the Ekman time (proportional to $E^{-1/2}$) emerges for $E \ll 1$.
However, whereas a viscous fluid is dissipative and results in relaxation, the superfluid is dissipation-less and oscillates.
For $E_s=0.01$ we find $\Omega t_P=8.35$ in agreement with Figure \ref{fig1}.

The weak pinning limit, $\gamma \ll \left[4\left(1+K\right)k\right]^{-1/2}$, (\ref{eq3.22}) gives
\begin{equation}
 s_+= -\frac{2}{k \gamma} \,, \hspace{5mm} s_- =-2 (1+K)\gamma\,. \label{eq3.27}
\end{equation}
Because $s_- \ll s_+$ in this limit, (\ref{eq3.23}) becomes
\begin{equation}
 f=\frac{1}{1+K}\left(1+K e^{s_-t}\right)\,. \label{eq3.28}
\end{equation}
This exponential decay is a result of the friction force of the vortex lines as they slide on the boundaries.
When $K=0$ the timescale for the relaxation is
\begin{equation}
  t_R=\frac{1}{\Omega |s_-| }=\frac{1}{2 \Omega \gamma }\,. \label{eq3.29}
\end{equation}
which is the result obtained in \citet{rei93}. 
Because the weak pinning limit was studied by \citet{rei93}, the principal focus of this paper will be on our new result in the perfect pinning limit.

\subsection{Ekman pumping} \label{sec3.2}

To visualize the flow for $E_s \ll 1$ we can invert the other variables in this limit.  We obtain
\begin{eqnarray}
  V_s&=&V_s^I+V_s^B \,, \label{eq3.30} \\
 \chi_s &=&\chi_s^I +\chi_s^B\,. \label{eq3.31}
\end{eqnarray}
where we have separated the solution into the interior flow and boundary layer corrections (denoted with subscripts $I$ and $B$ respectively) given by,
\begin{eqnarray}
V_s^I&=&\frac{1}{1+K}\left[1-\left(\frac{s_+ e^{s_- t} -s_- e^{s_+ t}}{s_+-s_-} \right) \right] \label{eq3.32} \\
V_s^B&=&\frac{1}{1+K}\left(\frac{s_+ e^{s_+ t} -s_- e^{s_- t}}{s_+-s_-} \right)\left[e^{k(z-1)}-e^{-k(z+1)} \right] \,, \label{eq3.33} \\
 \chi_s^I &=&\frac{2 z \left(e^{s_- t}-e^{s_+ t} \right)}{k\left(s_+-s_-\right)} \,, \label{eq3.34} \\
 \chi_s^B &=&-\frac{2\left(e^{s_- t}-e^{s_+ t} \right)}{k\left(s_+-s_-\right)} \left[e^{k(z-1)}-e^{-k(z+1)}\right]\,, \label{eq3.35} 
\end{eqnarray}
There is no boundary layer component for the pressure, which is given by
\begin{eqnarray}
  P_s&=&\frac{2}{1+K}\left[1-\left(\frac{s_+ e^{s_-t}-s_- e^{s_+ t}}{s_+-s_-}\right) \right]\,. \label{eq3.36}
\end{eqnarray}

Equations (\ref{eq3.32})--(\ref{eq3.36}) are akin to Ekman pumping in a viscous fluid \citep{gre63,ben74,van13}, where the exponentially decaying term has been replaced with the oscillatory  $e^{s_\pm t}$ terms. The variable $\chi_s$ is like a stream-function for a secondary flow [see (\ref{eq2.23})] which draws fluid radially inwards in the interior, delivers it to the boundary layer where it is pumped radially outwards.  However, when there is perfect pinning, the flow is oscillatory, with period given by (\ref{eq3.26}).  The Ekman circulation initially proceeds in the classical manner, clockwise in the region $0<z<1$ and counter clockwise in the region $-1<z<0$, but is periodically reversed.  This reversal is induced by vortex array.  In the boundary layer, the vortex array is sheared, storing potential energy in the form of vortex line tension.  The potential energy increases until the Ekman pumping is eventually slowed, halted and then reversed.  As there is no dissipation in the system, this energy exchange continues indefinitely.  
Because the flow obeys the Taylor-Proudman theorem to leading order, the interior azimuthal flow is columnar, as in classical Ekman pumping.  The vortices in the interior remain straight and move inwards and outwards with the secondary radial flow, increasing and decreasing the angular momentum of the fluid.

\begin{figure*}
  \includegraphics[width=0.8\textwidth]{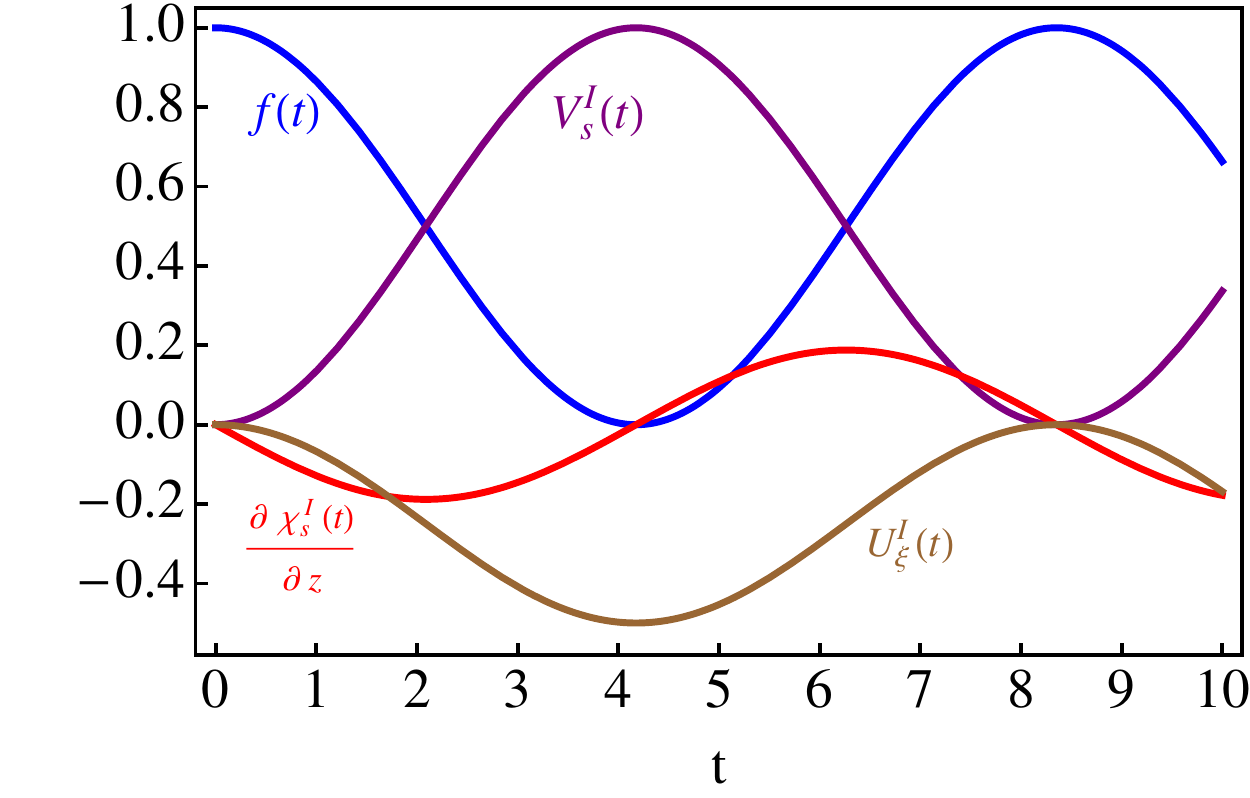}
\caption{Dimensionless angular velocity $f(t)$ (blue curve), azimuthal angular velocity, $V^I_s(t)$ (purple curve) and radial velocity (divided by $r$) $\partial \chi^I_s/\partial z$ (red curve) and the radial vortex line displacement $U_\xi$ (brown curve).  
We take $E_s=0.01$, $K=1$ and $\gamma\rightarrow\infty$.
Angular momentum is exchanged between the azimuthal flow and the container (blue and purple curves).  The radial flow (red curve) is $\pi/2$ radians out of phase with the azimuthal angular velocity. The vortices (brown curve) move inwards and outwards with the radial flow.}
\label{fig3}    
\end{figure*}

In Figure \ref{fig3} we plot the dimensionless angular velocity of the container, $f(t)$ (blue curve), angular velocity of the fluid, $V_s^I(t)$ (purple curve) and radial velocity of the fluid (divided by $r$), $\partial \chi^I_s/\partial z$ (red curve) and the radial vortex line displacement $U_\xi$ (brown curve) for $E_s=0.01$, $K=1$ and $\gamma\rightarrow\infty$.  The top two curves (blue and purple) show that the angular velocity of the container and the interior fluid are sinusoidal and $\pi$ radians out of phase.
Therefore angular momentum is conserved between fluid and container, as required.
The bottom curves show that the radial component of the secondary flow and the radial vortex line displacement are $\pi/2$ radians out of phase.  As the radial flow shears the vortices in the boundary layer, the restoring force from the line tension increases until the radial flow is halted, and then reversed. At this turning point in the radial flow the vortex displacement is at a maximum. 
The vortices are only displaced radially inwards, because the azimuthal flow velocity is always greater than or equal to its initial value.

\begin{figure*}
  \includegraphics[width=0.5\textwidth]{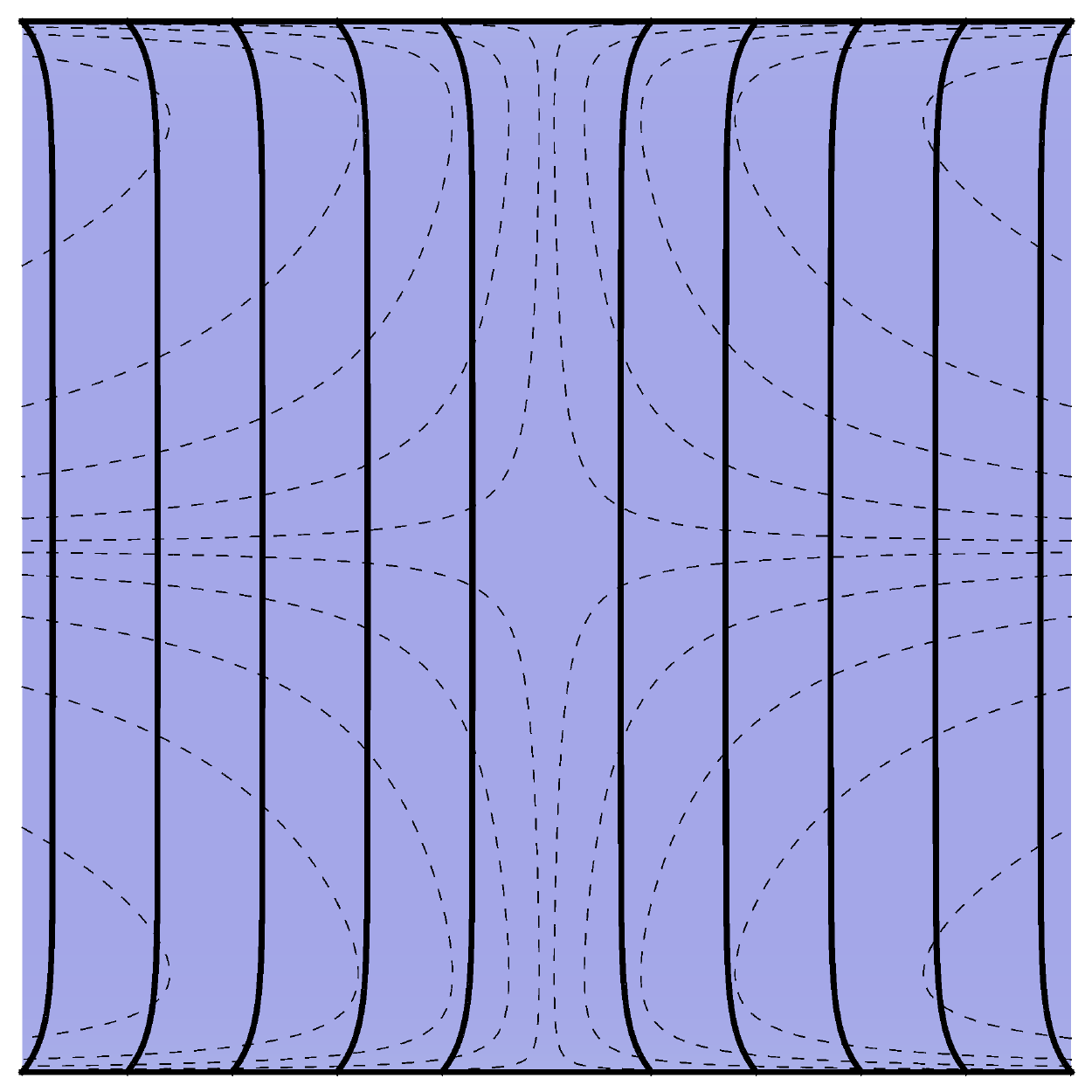}
  \includegraphics[width=0.5\textwidth]{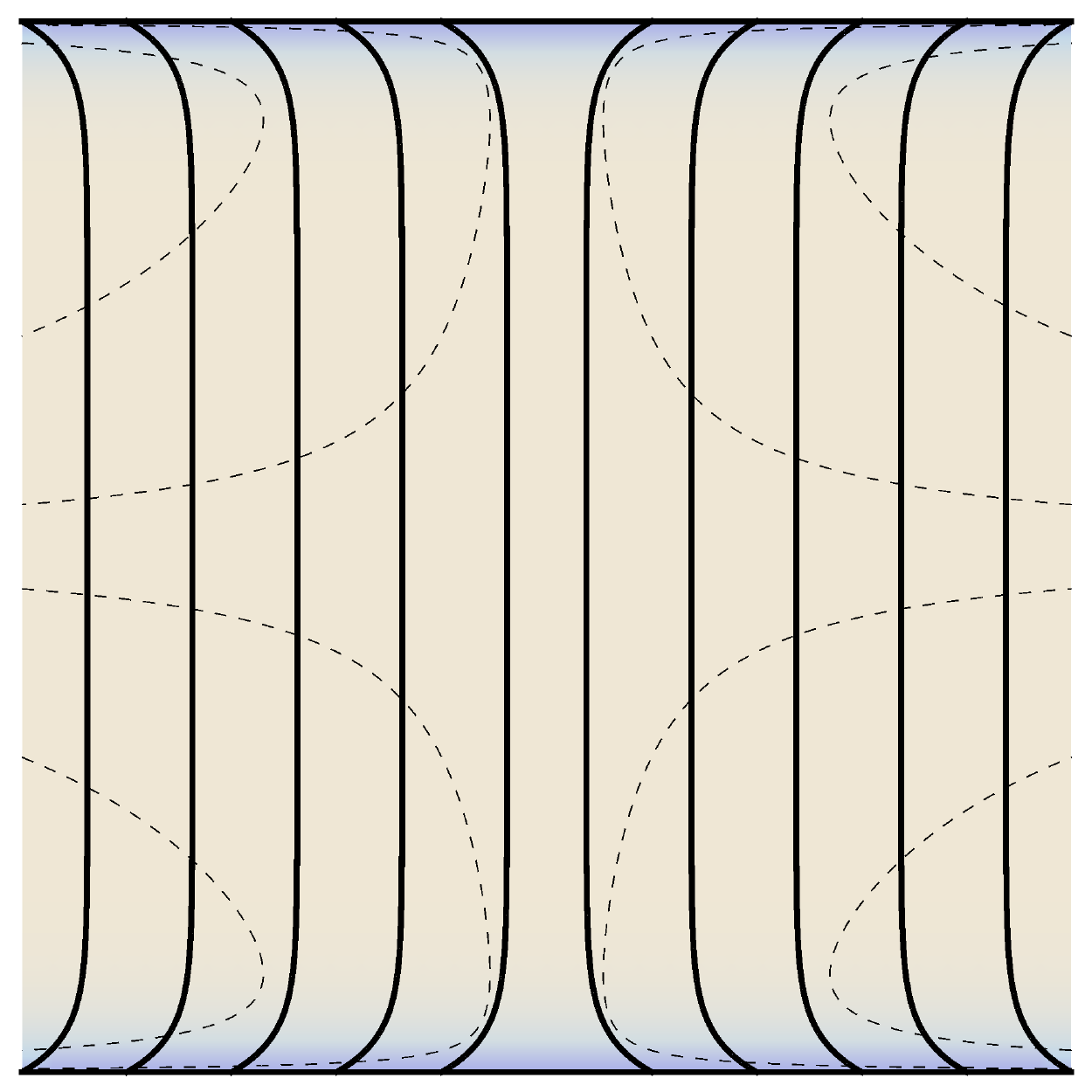}
\caption{Flow plots for $E_s=0.01$. Left: $t=2$ (near secondary flow maximum), right: $t=4$ (near secondary flow maximum).  
The color shading represents the magntude of the azimuthal angular velocity, $V_s^I(t)$, where blue zero and white is unity (c.f. Figure \ref{fig3}).  The dashed contours show streamlines for the secondary flow $\chi_s^I(t)$ with values: $\pm$ 0.002, 0.01, 0.025, 0.05, 0.1.  The solid lines show the vortex line displacement $U_\xi^I$, plotted with exaggerated amplitude for visibility.}
\label{fig4}    
\end{figure*}

Snapshots of the flow described by Figure \ref{fig3} are presented in Figure \ref{fig4} at the times $t=2$ (left panel) and $t=4$ (right panel) 
The magnitude of the fluid angular velocity $V_s$ is shown in color, where dark is zero and light is unity.  
The radial vortex line displacement $U_\xi$ is over-plotted as thick-black lines, scaled appropriately for illustrative purposes.
Contours for the stream-function of the secondary flow, $\chi_s$  are plotted as dashed lines for values of $\pm$ 0.002, 0.01, 0.025, 0.05, 0.1.  In the left panel ($t=2$) the secondary flow is near a maximum, represented by the large density of contours. 
The flow is clockwise in the upper-right quadrant and the secondary flow is moving the vortex lines inwards in the interior.
The angular velocity of the container and azimuthal flow are matched, so there is no boundary layer correction. 
In the right panel ($t=4$), the secondary flow is near a minimum.
The radial component of the secondary flow is turning from inwards to outwards in the interior.  The vortex lines are at their maximum extension, and will subsequently straighten to drive the outward radial flow in the interior.
The difference between the angular velocities of the container and fluid is greatest, and the boundary layer in the  azimuthal velocity is prominent.  


To make a final connection to Ekman pumping, we examine the governing equations in the limit $E_s\ll1$.  In the interior, we find from (\ref{eq3.32}) that the azimuthal velocity evolves on the timescale $E_s^{-1/4}\Omega^{-1}$ and from (\ref{eq3.34}) the stream-function for the secondary flow, $\chi_s$, scales as $E_s^{1/4}$. 
Applying these scalings, we find in exactly the same manner as for classical Ekman pumping [see \citet{gre63}, \S 5] the solutions for the interior flow satisfy
\begin{eqnarray}
 0 &=& 2 V_s^I -P_s \,, \label{eq3.37}\\
 0 &=& \frac{\partial V_s^I}{\partial t}+2 \frac{\partial \chi_s^I}{\partial z} \,. \label{eq3.38}
\end{eqnarray}
The second, (\ref{eq3.38}) describes the geostrophic balance in the interior.
In the boundary layer we have
\begin{eqnarray}
  0&=&\left(2-E_s \frac{\partial^2}{\partial z^2}\right)V_s^B \,, \label{eq3.39}\\
  0&=&\left(2-E_s \frac{\partial^2}{\partial z^2}\right) \frac{ \partial \chi_s^B}{\partial z} \,.\label{eq3.40}
\end{eqnarray}
The boundary conditions are given by (\ref{eq2.37})--(\ref{eq2.39}), where (\ref{eq2.38}) is replaced by
\begin{equation}
  \frac{\partial }{\partial t}  \frac{\partial \chi_s^B}{\partial z}-2 f \pm  \frac{E_s}{\gamma} \frac{\partial^2 \chi_s^B}{\partial z^2}+ P_s =0 \,, \label{eq3.41}
\end{equation}
at $z=\pm 1$.
Finally, the initial conditions are
\begin{eqnarray}
 V_s^I(0)=0 \,, \label{eq3.42}\\
 f(0)=0 \,, \label{eq3.43}\\
 \chi_s^I(0)=0\,,\label{eq3.44}
\end{eqnarray}
Equations (\ref{eq3.37})--(\ref{eq3.44}) are sufficient to recover the approximate solution (\ref{eq3.30})--(\ref{eq3.36}).  They are the equivalent of the boundary layer approximation derived for classic spin-up by \citet{gre63}.  They will be used in \S\ref{sec4} to derive a general solution for the two-fluid system.

The mechanism described above possesses many properties similar to Ekman pumping.  In particular, boundary layers form on boundaries orthogonal to the rotation axis, driving a radial outflow.  This flow is replenished by fluid from the interior, generating a secondary circulation.  However, there are important differences.  Classical Ekman pumping is dissipative, while the mechanism described here is dissipation-less.  Also, in classical Ekman pumping, the Coriolis force couples the azimuthal and radial velocity in the boundary layer, driving the secondary flow.  Here, the velocity components in the boundary layer are uncoupled [see (\ref{eq3.39}) and (\ref{eq3.40})], but
it is the boundary condition itself (\ref{eq2.20}) that couples the velocity components.  

Finally, (\ref{eq3.37}) -- (\ref{eq3.41}) can be solved to obtain a boundary condition for the interior flow analogous to \citet{gre68} [see (2.6.13)] that relates the differential vorticity at the boundary with the secondary inflow into the boundary layer,
\begin{equation}
 \left( \frac{1}{\sqrt{2 \Omega \nu_s}}\frac{\partial }{\partial t}\mp \frac{1}{L\gamma} \right) \hat{k} \cdot \boldsymbol{v}^I= \hat{k} \cdot  \nabla \times \left( \boldsymbol{v}^I-\boldsymbol{v}_B \right)\,,
\end{equation}
at $z=\pm 1$, where $\boldsymbol{v}^I$ denotes the velocity field in the interior.
This relation applies at the wall in cylindrical geometry and applies only to a single fluid.

\section{Helium II}\label{sec4}

\subsection{Experiments} \label{sec4.1}

The response of a superfluid-filled container following an impulsive acceleration has been investigated in series of experiments conducted by \citet{tsa80}.  
In \S 5 of \citet{tsa80}, plexiglass cylinders are coated with a powder to increase vortex adhesion, and oscillatory motion of the container is reported.  
The experiment is conducted at $1.52\,{\rm K}$ with a rotational frequency of $3\,{\rm rad\,s^{-1}}$.
The vessel is $64\,{\rm mm}$ in diameter, $50\,{\rm mm}$ in height and $0.2\,{\rm mm}$ thick.
The response of the vessel is given in  Figure 27c of \cite{tsa80}.  The magnitude of the angular velocity after the spin up is not given, but appears to be around $3.6\,{\rm rad\,s^{-1}}$.  The response comprises an exponential decay of $0.3\,{\rm rad\,s^{-1}}$ over approximately $40\,{\rm s}$, followed by steady oscillation with a period of $40\,{\rm s}$, decaying over a timescale of roughly $480\,{\rm s}$.  The oscillation period also appears to decrease with time, probably a result of non-linear effects as the experiment has Rossby number $\epsilon \sim (3.6-3)/3=0.2$.

\citet{tsa80} also report the following effects.  The oscillation period is unaffected by the radius of the cylinder.  This was tested by inserting coaxial cylinders of both $22\,{\rm mm}$ and $43\,{\rm mm}$ inside the main cylinder.  However, the frequency was observed to double when the container height was halved, which was achieved by inserting disks into the cylinder.  This behavior is qualitatively consistent with the columnar flow of Ekman pumping and the solution presented in \S\ref{sec3}.

To apply the our theory to the Tsakadze experiments, we need to generalize the solution in \S\ref{sec3} to the two-fluid system in \S\ref{sec2}.   For the experiment described above, the dimensionless parameters at $1.52\,{\rm K}$ are \citep{don98}
\begin{eqnarray}
 E &=&4.07\times 10^{-5} \left( \frac{\nu_n}{7.64\times 10^{-4}\, {\rm cm^2\,s^{-1}}} \right) \left( \frac{\Omega}{3\,{\rm rad\,s^{-1}}} \right)^{-1} \left(\frac{L}{2.5\,{\rm cm}}\right)^{-2}\,,   \nonumber \\
 E_s &=&5.95\times 10^{-5} \left( \frac{\nu_s}{1.12\times 10^{-3}\, {\rm cm^2\,s^{-1}}} \right) \left( \frac{\Omega}{3\,{\rm rad\,s^{-1}}} \right)^{-1} \left(\frac{L}{2.5\,{\rm cm}}\right)^{-2}\,, \nonumber \\
{\rm Ma} &=& 3.2 \times 10^{-4}\left( \frac{u_1}{2.35\times 10^4\, {\rm cm\,s^{-1}}} \right)^{-1} \left( \frac{\Omega}{3\,{\rm rad\,s^{-1}}} \right) \left(\frac{L}{2.5\,{\rm cm}}\right) \,, \nonumber
\end{eqnarray}
\begin{eqnarray}
 B&=& 1.2748\,, \nonumber\\
 B'&=& 0.2922\,, \nonumber\\
 \rho_n&=&0.1206 \rho \label{eq4.1}\,.
\end{eqnarray} 
The third parameter, ${\rm Ma}$, where $u_1$ is the first sound velocity, is the Mach number and indicates the relative importance of compressibility effects.  Clearly for helium II the incompressibility approximation is valid.
For the experiment described above, all the parameters of our theory are given by (\ref{eq4.1}) except $K$.  In experiments involving spheres and no pinning, \citet{tsa80} present the steady-state spin down of empty and filled vessels from which one can determine $K\sim 0.8$ \citep{van11a}. Unfortunately, \citet{tsa80} do not provide enough data to determine $K$ for this experiment, and we take $K=1$.

\subsection{Two-fluid solution} \label{sec4.2}

Solving the full system of equations in \S\ref{sec2} poses a formidable challenge.  In helium II, we have $E \ll 1$ and $E_s \ll 1$ but we still have $E_s \sim E$, so that neither viscosity or vortex tension is negligible.  Also, because $B,B'\sim 1$, the mutual friction coupling time between the two fluid components is of the order of the rotation period, and the normal fluid and superfluid are approximately locked together over the Ekman time \citep{rei93,van13,van14a}.  

To solve the two-fluid system we use a boundary layer approximation like that used in \S 5 of \citet{gre63} and presented in \S\ref{sec3.3}.  
We assume that the azimuthal velocity is symmetric about the $z=0$ plane, and hence $\chi_{n,s}$ are antisymmetric.
In the interior we have
\begin{eqnarray}
0&=& 2 V^I_n- \frac{\rho_s B'}{\rho}\left(V^I_n-V^I_s\right)  - P_n \, , \label{eq4.2} \\
0&=&\frac{\partial V^I_n}{\partial t}+ 2\frac{\partial \chi^I_n}{\partial z} + \frac{\rho_s B}{\rho}\left(V^I_n-V^I_s\right) -\frac{\rho_s B'}{\rho} \left( \frac{\partial \chi^I_n}{\partial z} - \frac{\partial \chi^I_s}{\partial z} \right) \, ,  \label{eq4.3} 
\end{eqnarray}
\begin{eqnarray}
0&=&2 V^I_s +\frac{\rho_n B'}{\rho} \left(V^I_n- V^I_s\right) - P_s \, , \label{eq4.4}\\
0&=& \frac{\partial V^I_s}{\partial t} +2  \frac{\partial \chi^I_s}{\partial z}-\frac{\rho_n B}{\rho}\left(V^I_n- V^I_s\right) + \frac{\rho_n B'}{\rho}   \left( \frac{\partial \chi^I_n}{\partial z}- \frac{\partial \chi^I_s}{\partial z}\right) \, , \label{eq4.5} 
\end{eqnarray}
while in the upper boundary layer we have
\begin{eqnarray}
0&=&\left( -  E \frac{\partial^2}{\partial z^2} +\frac{\rho_s B}{\rho} \right) \frac{\partial \chi^B_n}{\partial z}  - \left(2-\frac{\rho_s B'}{\rho} \right) V^B_n \nonumber \\
&& -\frac{\rho_s B}{2\rho}\left(2 -E_s\frac{\partial^2}{\partial z^2}\right) \frac{\partial \chi^B_s}{\partial z}-\frac{\rho_s B'}{2\rho}\left(2-E_s\frac{\partial^2}{\partial z^2}\right)V^B_s  \, , \label{eq4.6} \\
0&=&\left(  - E \frac{\partial^2}{\partial z^2}+ \frac{\rho_s B}{\rho} \right)V^B_n + \left(2 -\frac{\rho_s B'}{\rho} \right)\frac{\partial \chi^B_n}{\partial z} \nonumber \\
&& -\frac{\rho_s B}{ 2 \rho} \left( 2 - E_s \frac{\partial^2}{\partial z^2}\right)V^B_s + \frac{\rho_s B'}{2 \rho} \left( 2 - E_s \frac{\partial^2}{\partial z^2}\right)\frac{\partial \chi^B_s}{\partial z}\, ,  \label{eq4.7} 
\end{eqnarray}
\begin{eqnarray}
0&=&\frac{\rho_n B}{2 \rho} \left(2 - E_s \frac{\partial^2}{\partial z^2}  \right)\frac{\partial \chi^B_s}{\partial z} - \frac{\rho_n B}{\rho} \frac{\partial \chi^B_n}{\partial z}\nonumber \\
&&- \left(1-\frac{\rho_n B'}{2\rho} \right) \left(2 - E_s \frac{\partial^2 }{\partial z^2}\right) V^B_s -\frac{\rho_n B'}{\rho}  V^B_n  \, , \label{eq4.8}\\
0&=&\frac{\rho_n B}{2\rho}\left(2-E_s \frac{\partial^2 }{\partial z^2} \right) V^B_s -\frac{\rho_n B}{\rho} V^B_n \nonumber \\
&&+\left(1 -\frac{\rho_n B' }{2\rho} \right) \left(2-E_s \frac{\partial^2}{\partial z^2}\right)  \frac{\partial \chi^B_s}{\partial z} + \frac{\rho_n  B' }{\rho} \frac{\partial \chi^B_n}{\partial z} \, . \label{eq4.9} 
\end{eqnarray}
These equations reduce to  (5.3)--(5.10) in \citet{gre63} and (\ref{eq3.37})--(\ref{eq3.40}) in the present paper when $B=B'=0$.

The governing equation for the motion of the container is (\ref{eq2.32}).
The boundary conditions for the superfluid are (\ref{eq2.37}), (\ref{eq3.41}) and (\ref{eq2.39}).  For the normal fluid, the required boundary conditions are [\citet{gre63}, \S 5]
\begin{eqnarray}
  V_n-f&=&0 \,, \label{eq4.10} \\
  \frac{\partial \chi_n^B}{\partial z}&=&0 \,, \label{eq4.11} \\
  \chi_n&=&0\,.  \label{eq4.12}
\end{eqnarray}
The initial conditions are (\ref{eq3.42})--(\ref{eq3.44}) for the superfluid and 
\begin{equation}
  V_n^I(0)=0\,, \label{eq4.13}
\end{equation}
for the normal fluid.
The general solution is presented in \S\ref{secA1}. 
Because of its complexity, we do not present the final result in algebraic form.
For the numbers (\ref{eq4.1}) and taking $K=1$, $\gamma \rightarrow \infty$ and $\Omega=3\,{\rm rad\,s^{-1}}$, we obtain
\begin{eqnarray}
  f(t)&=& 0.500000+0.000023e^{-3.82908 t} \nonumber \\
 &+&e^{-0.004182 t}\left[0.499977 \cos (0.58827 t)+0.001126\sin(0.58827 t)\right]\,, \label{eq4.14}
\end{eqnarray}
where $t$ is in seconds.
The first term is the steady-state term, which corresponds to the centre-of-mass of the system at $1/(1+K)$.  The second term corresponds to a rapid damping and has low amplitude.  This term represents the strong mutual friction between the two components.  The third term is an oscillation, weakly damped by viscosity.  It has a period of $10.7\,{\rm s}$ and a damping time $239.1\,{\rm s}$.

The result (\ref{eq4.14}) suggests that viscosity and mutual friction have little effect on the final result.  This is because the Ekman time and secondary flow scale as $E^{1/2}$ in classical Ekman pumping, but scale as $E_s^{1/4}$ for the oscillatory superfluid Ekman pumping mechanism identified in this paper.  
For the numbers in (\ref{eq4.1}), we find
\begin{equation}
  E_s^{1/4}=0.088\,, \hspace{5mm}  E^{1/2}=0.006\,. \label{eq4.15}
\end{equation}
Therefore the superfluid mechanism is dominant.  Neglecting viscosity ($E=0$) and assuming strong mutual friction coupling ($B,B'\sim 1$), an approximate solution can be obtained.  The details are presented in \S\ref{secA2}.  
The result has the form (\ref{eq3.23}), with $s\pm$ now given by
\begin{equation}
  s_\pm=-\frac{ 1}{k \gamma }\left[ 1\pm \sqrt{1-\frac{4\left(1+K\right)k \gamma^2 \rho_s}{ \rho }} \right]\,. \label{eq4.16}
\end{equation}
The oscillatory Ekman pumping mechanism identified in \S\ref{sec3} is operating in the superfluid, with the normal fluid locked to it as a result of the strong mutual friction coupling.  This mass loading of the superfluid results in a factor of $\rho_s/\rho$ in the spin-up time.
Only the leading order normal fluid velocity is involved in the oscillation; there is no secondary flow in the normal fluid.  The reader is referred to \S\ref{secA2.2} for the solution for all variables in this limit.

\begin{figure*}
  \includegraphics[width=0.75\textwidth]{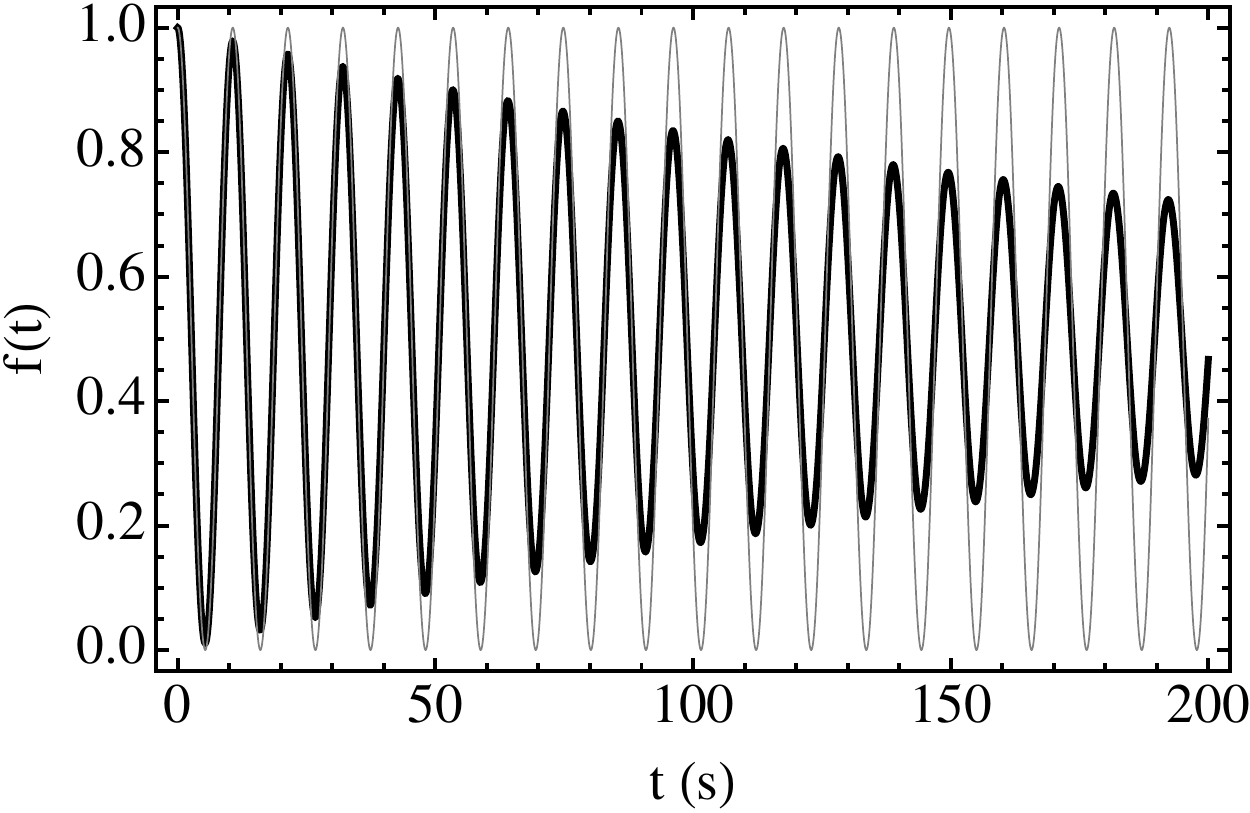}
\caption{Comparison of the general solution [(\ref{eq4.14}), thick black curve] and the approximate solution [(\ref{eq3.23}) and (\ref{eq4.16}), thin gray curve] assuming zero viscosity ($E=0$) and strong mutual friction couping ($B,B' \sim 1$).  
Viscosity has a negligble effect on the period, which is the same for both solutions.  
The experimentally observed period decays from an initial maximum of $40\,{\rm s}$.  \label{fig5}   }
 \end{figure*}

A comparison between the general solution (\ref{eq4.14}) and the approximate solution  [(\ref{eq3.23}) and (\ref{eq4.16})] assuming $E=0$ and $B,B'\sim1$ is shown in Figure \ref{fig5}.  The period of the approximate solution is identical to the general result, suggesting it depends weakly on viscosity.  The approximate solution is dissipation-less, whereas the general solution is damped on a timescale of a few hundred seconds.
When $\gamma\rightarrow\infty$, the period of oscillation at $T=1.52\,{\rm K}$ is
\begin{equation} \label{eq4.17}
  t_P=10.7\,{\rm s} \left(\frac{3\,{\rm rad\,s^{-1}}}{\Omega}\right)^{-1} \left(\frac{5.95\times 10^{-5}}{E_s}\right)^{-1/4} \left(\frac{0.8794}{\rho_s/\rho}\right)^{-1/2} \left(\frac{2}{1+K}\right)^{-1/2} \,.
\end{equation}
This is shorter than the maximum observed period of oscillation in the Tzakadze experiments, which decays from an initial maximum of approximately $40\,{\rm s}$.   
The period can be lengthened by decreasing the frictional pinning force (decreasing $\gamma$), however, this quickly shortens the damping time to much less than that observed in the experiments. 
Because of the precision to which the parameters in (\ref{eq4.1}) have been measured, the discrepancy must arise either because the linear hydrodynamic theory is invalid, or one or more of the experimental parameters have been erroneously reported.  
The former is certainly likely; as discussed in \S\ref{sec4.1}, the Rossby number is $0.2$, so non-linear effects may be present.  The observed decay of the oscillation period cannot be captured by the linear theory.  The impulsive spin-up of the container cannot be done instantaneously as required for a perfect comparison with theory, and this may explain this initial rapid decay of the container motion observed in the experiments.
Errors in reporting the experimental parameters is also possible, because \citet{tsa80} report some  inconsistently (e.g., vessel radii), and others not at all (e.g., the vessel moment of inertia, which we have assumed equal to that of the fluid).  A thorough comparison between theory and experiment will require new experiments.  Such experiments could be combined with the well known properties of helium II to provide a useful test of both the HVBK theory and the calculations in this paper, as in \citet{van11a}.  Alternatively, the container response predicted here may be a useful check of known superfluid parameters in helium II, or used to constrain parameters in condensates where they remain unknown \citep{van11b}.

\subsection{Tkachenko oscillations} \label{sec3.3}

In our calculations, we have solved the complete equations of \citet{cha86}, in order to apply the most sophisticated hydrodynamic theory available to the helium II experiments.
Although this theory includes the extra term (\ref{eq2.9}) over the HVBK theory, we find that the contribution from this force vanishes.  
This result was also obtained by \citet{rei93}.
The reason for this lies in the ansatz (\ref{eq2.22}), where the flow is assumed to be axisymmetric and the radial dependence of azimuthal velocity matches the boundary conditions.  
The equation of motion for the vortex lines requires that $\boldsymbol{\xi}$ has the same radial dependence as the superfluid velocity, and upon substituting (\ref{eq2.22}) into (\ref{eq2.9}), we find that $\boldsymbol{\sigma}=0$.

Physically, (\ref{eq2.9}) describes the restoring force experienced by the vortex array as a result of displacements of vortices from their equilibrium configuration. Hence, the vortex array experiences a restoring force in response to shear deformations, giving rise to Tkachenko oscillations.  However, in the spin-up problem considered here, the top and bottom boundaries are rigid bodies and the impulsive acceleration of the boundary does not displace any vortices from their equilibrium configuration. Therefore we do not necessarily expect Tkachenko modes to be excited in the experiments of \citet{tsa80}.

Tkachenko modes may be excited at the cylinder wall, which has been neglected in our analysis.  
However the required boundary conditions required at the cylinder wall are not discussed by \citet{bay83}. 
It is known that it is energetically favorable for vortex-free region to form adjacent to the cylinder wall, making it difficult for changes in the angular velocity of the side-wall to be communicated to vortex array.  
Even if they were excited, Tkachenko oscillations on timescales of order $E_s^{-1}\Omega^{-1}$, much less than the Ekman pumping mechanism described above, which has timescale $E_s^{-1/4}\Omega^{-1}$.  
Therefore the motion of the vortex lines is determined by the vortex tension producing Ekman pumping, and the restoring force from the lattice deformations is too slow to have any effect.

Therefore,  Tkachenko oscillations are unlikely to be excited in the experiments of \citet{tsa80}.  
Tkachenko oscillations have been imaged in rapidly-rotating dilute-gas Bose-Einstein condensates, excited by blasting atoms or creating a dip in the trapping potential at the centre of the condensate \citep{cod03}.  Here, the experimental setup and excitation mechanism are appropriate to excite Tkachenko modes, whereas in the experiments of \citet{tsa80} they are not.
Therefore, the Ekman pumping mechanism presented here provides a much more natural explanation for the oscillations observed in the \citet{tsa80} experiments.

\section{Neutron stars} \label{sec5}

Neutron stars are compact stellar corpses formed by the core collapse of a massive star after a supernova.  The resulting compact object typically has a radius of 12-14 km and a mass of 1.4-2 times that of the sun.  The outer kilometer of the star is a rigid, highly conducting crustal lattice interspersed with a neutron superfluid, while the core comprises approximately 95\% superfluid neutrons, 5\% electrons and superconducting protons, and a small fraction of muons.  In the deep core, the composition remains uncertain.

Neutron stars are also endowed with strong magnetic fields, ranging from $10^8$--$10^{15}$ Gauss.  In isolated radio pulsars, radio waves are beamed along the magnetic axis and create a lighthouse effect that is observed as pulsations on Earth.  Occasionally, these objects are observed to `glitch', where the rotational frequency of the star suddenly increases by up to one part in $10^5$ \citep{esp11}.  The distribution of glitch sizes and waiting times is consistent with the hypothesis that they arise from a self-organised, collective, avalanche process \citep{and75b,mel08,war13}.
During a glitch, angular momentum stored in the decoupled superfluid components of the neutron star is transferred erratically to the solid crust ($\sim1\%$ of the total moment of inertia) \citep{lin92}, although the exact source of this angular momentum is uncertain \citep{and12,cha13}.
The trigger is often ascribed to collective unpinning of quantized superfluid vortices, induced when crust-superfluid differential rotation (and hence the Magnus force) exceeds a threshold \citep{bay92,war12}.

The glitch event is followed by a recovery phase, during which the components inside the star adjust to a new equilibrium configuration \citep{bay69a}.  This recovery is typically dominated by a step increase in the rotational frequency of the star, a step change in the frequency derivative, and a quasi-exponential relaxation \citep{van10}.  
However, quasi-exponential oscillations have also been reported following some glitches.
A post-glitch oscillation with a period of approximately four months was reported in the Crab pulsar soon after its discovery \citep{rud70a,rud70b}.  The 1975 and 1986 Crab glitches were observed to ``overshoot'' their final rotation frequency during their recovery \citep{won01,van10}, which could be potentially be explained by a damped oscillation with a period of approximately 200-300 days.
Oscillations also appear to be present in timing residuals following the 1988 Christmas Vela glitch [see figure 2 of \citet{mcc90}], which, if real, appear to have a period of $\sim 20\, {\rm days}$.
These oscillations have received little attention observationally or theoretically, but may shed light on the interior of neutron stars.  Here we consider the possibility that the superfluid oscillations described in the previous sections may be operating in neutron stars, and ask what the observable consequences are for glitch recovery. 

During glitch recovery, the proton-electron plasma in the core is coupled to the highly conducting crust via the magnetic field on a time-scale of seconds \citep{eas79a,van14b}, and hence, these components are assumed to be rigidly locked together.
The superfluid neutrons are usually assumed to respond changes in the angular velocity of the crust via the interaction of the neutron vortices with the proton-electron plasma in the core \citep{alp84,men91a,men91b,gla11}.  The electrons scatter from the neutron vortices that are magnetized by the entrainment of proton currents, giving the mutual friction force 
\begin{eqnarray}
 \boldsymbol{F}&=&-\frac{\rho}{\rho_n} \beta_0 \hat{\omega}_s\times\left[\boldsymbol{\omega}_s\times\left(\boldsymbol{v}_L -\boldsymbol{v}_n\right)\right] \,, \label{eq5.1}
\end{eqnarray}
where the subscript $n$ refers the the proton-electron plasma, which plays the role of the normal fluid component and $s$ refers to the neutron superfluid.
The coefficient $\beta_0$ is given by \citet{men91b}
\begin{equation}
  \beta_0=1.1\times10^{-2}\frac{\left(y-1\right)^2 x^{7/6}}{y^{1/2}\left(1-x\right)}\,, \label{eq5.2}
\end{equation}
where $x$ is the proton fraction ($\rho_n/\rho$ in the notation used here) and  $y$ is the normalized effective mass of the proton from Fermi liquid theory. 
The coefficient $\beta'_0$ that appears in (\ref{eq5.1}) in HVBK theory is typically taken as zero in neutron stars.
The motion of magnetized neutron vortices is also inhibited by their pinning to flux tubes.
When a vortex is pinned it moves with the normal fluid so that $\boldsymbol{v}_L=\boldsymbol{v}_n$.  However, an unpinned vortex moves with velocity given by (\ref{eq2.12}).
Solving the vortex line equation of motion (\ref{eq2.12}) gives
\begin{eqnarray}
 \boldsymbol{v}_L&=&\boldsymbol{v}_n-\frac{1}{1+\beta_0^2}\left\{\hat{\omega}_s\times\left[\hat{\omega}_s\times\left(\boldsymbol{v}_s-\boldsymbol{v}_n\right)+\boldsymbol{\sigma}+\boldsymbol{t}\right] \right.\nonumber  \\
&+&\left. \beta_0 \left[\hat{\omega}_s\times\left(\boldsymbol{v}_s-\boldsymbol{v}_n\right)+\boldsymbol{\sigma}+\boldsymbol{t}\right] \right\} \,.\label{eq5.3}
\end{eqnarray}
At zero temperature, all vortices are pinned to flux tubes.  At finite temperature a vortex can become thermally excited with activation energy $A$ upon unpinning.  
The activation energy for unpinning depends on the magnetic energy between a vortex and a flux tube $E_p$, the dimensionless vortex tension $\mathcal{T}$, the angular velocity lag $\Delta\omega$ and the critical angular velocity lag for unpinning $\Delta\omega_c$ as
\begin{equation}
  A=5.1 E_p \mathcal{T}^{1/2}\left(1-\frac{\Delta\omega}{\Delta\omega_c}\right)^{5/4}\,. \label{eq5.5a}
\end{equation}
Therefore the mutual friction force has a non-linear dependence on the velocity through $\Delta\omega$.
However, to simplify our analysis we take $A$ as constant, noting that a rigorous calculation should include this non-linear dependence.

The partition function for this two-state system is
\begin{equation}
  Z= 1+ e^{-\beta A}\,,
\end{equation}
where $\beta=\left(k_B T\right)^{-1}$, $k_B$ is Boltzmann's constant and $T$ is the temperature.  When the thermal energy is much greater than the activation energy for unpinning a vortex  $\beta^{-1} \gg A$, the all vortices are thermally activated and move with average velocity given by (\ref{eq5.3}).  However, when the thermal energy is much less than the activation energy for unpinning $\beta^{-1} \ll A$, the vortex movement is exponentially suppressed pinning the vortices to the normal fluid.  
For slow vortex slippage, $e^{-\beta A} \ll 1$ and $Z\approx 1$ and the ensemble average vortex line velocity is given by
\begin{eqnarray}
 \langle \boldsymbol{v}_L \rangle_\beta -\boldsymbol{v}_n &=& \langle \boldsymbol{v}_L  - \boldsymbol{v}_n\rangle_\beta \approx Z^{-1} e^{-\beta A} \left(\boldsymbol{v}_L  -\boldsymbol{v}_n\right) \nonumber \\
&=&-\frac{1}{1+\beta_0^2}\left\{\hat{\omega}_s\times\left[\hat{\omega}_s\times\left(\boldsymbol{v}_s-\boldsymbol{v}_n\right)+\boldsymbol{\sigma}+\boldsymbol{t}\right] \right.\nonumber  \\
&+&\left. \beta_0 \left[\hat{\omega}_s\times\left(\boldsymbol{v}_s-\boldsymbol{v}_n\right)+\boldsymbol{\sigma}+\boldsymbol{t}\right] \right\} e^{- \beta A}\,. \label{eq5.3b}
\end{eqnarray}
Substituting the vortex line velocity from (\ref{eq5.3b}) into the vortex line conservation equation (\ref{eq2.10}) and integrating, we obtain an HVBK-like equation for the superfluid (\ref{eq2.2}) with modified mutual friction coefficients
\begin{eqnarray}
  B&=&\frac{2 \rho}{\rho_n}\frac{\beta_0}{1+\beta_0^2}e^{-\beta A}\,, \label{eq5.4}\\
B'&=&\frac{2 \rho}{\rho_n}\left[1-\frac{1}{1+\beta_0^2}e^{-\beta A}\right]\,. \label{eq5.5}
\end{eqnarray}
For a detailed derivation of the theory of vortex slippage, the reader is referred to \citet{lin14a}, where (\ref{eq5.3b}) and (\ref{eq5.5}) are rigorously derived [see (48) and (51) in that reference].

Before proceeding, we assess the validity of the incompressibility assumption for the present application.  In neutron stars the speed of sound is approximately one fifth of the speed of light (see e.g., \citet{rei92}), giving a Mach number $1.7\times 10^{-3}$.  Therefore the incompressibility assumption in neutron stars is valid. 

We can now apply the theory in \S\ref{sec2} to neutron stars.  
Because the normal fluid is rigidly locked to the crust via the magnetic field, we have $\boldsymbol{v}_n= \boldsymbol{\Omega}_c r$.  Therefore, it is a single-fluid problem involving only the superfluid and we can repeat the analysis in \S\ref{sec3} with the addition of mutual friction terms.  In this case, the crust plays the role of the container.  
We assume that the vortices are perfectly pinned to the crust with $\gamma \rightarrow \infty$ at $z=\pm 1$.
In the core of the star (the region $-1<z<1$), we expect superfluid oscillations as illustrated in \S\ref{sec3} where vortex lines are straight in the core, except a small boundary layer region adjacent to the crust.
However, the motion of the neutron superfluid in the core is now impeded by the interaction with the flux-tubes in the proton electron plasma, which is described by the mutual friction force.

For the superfluid oscillation mechanism to operate, vortices pinned in the crust must communicate with vortices in the core via vortex line tension.  In the outer crust, neutrons pair in a $^1 S_0$ state, while in the core they are expected to pair in a $^3 P_2$ state.  The $^1 S_0$ phase is expected to be lost near the base of the crust, however current calculations of the pairing gaps for the $^3 P_2$ phase are not certain enough to know at what depth superfluidity will be restored.   If the two phases are connected, then the relevant boundary condition at the interface are also unknown.  It may be that the vortices in the crust and core are not connected, in which case the superfluid oscillations described here will not occur.  

Repeating the analysis in \S\ref{sec3} with the mutual friction terms, the result (\ref{eq3.21}) generalizes to
\begin{eqnarray}
  \bar{\Delta}\approx \frac{ k e^{2 k}}{2 \rho s \left(\rho s+\rho_n B\right)}&&\left\{\left(2\rho-\rho_n B' \right)\left[ 2 K \rho_s+\left(2\rho-\rho_n B' \right) \right] \right. \nonumber  \\
&+&\left. \left(\rho s+\rho_n B\right)\left(\rho k s+\rho_n B\right)\right\}\,, \label{eq5.6}
\end{eqnarray}
and the inverse Laplace transform is 
\begin{equation}
 f(t)=\frac{1}{\left(2 \rho-\rho_n B'\right)+2 K \rho_s}\left[\left(2 \rho-\rho_n B'\right)+\left(\frac{2 K \rho_s}{s_+-s_-}\right)\left(s_+ e^{s_-t}-s_- e^{s_+ t}\right)\right]\,, \label{eq5.7}
\end{equation}
where
\begin{equation}
  s_\pm=-\frac{\rho_n B}{2 \rho}\pm \sqrt{\frac{k \left(\rho_n B\right)^2-4\left(2\rho-\rho_n B' \right)\left[ 2 K \rho_s+\left(2\rho-\rho_n B' \right) \right]}{4 \rho^2 k}}\,. \label{eq5.8}
\end{equation}
In a typical neutron star, we have $\rho_n/\rho=x\sim 0.05$, $\Omega=100\,{\rm rad\, s^{-1}}$, $0.3\leq y = m_p^*/m_p \leq 0.7$ and  $K\approx 50$ \citep{men91a,men91b,van10}.  The pinning fraction in the core is uncertain.  Hence we have $\beta_0\ll 1$ [see (\ref{eq5.2})], and also $K \rho_s \gg \rho$ because the neutron superfluid is the dominant contribution to the moment of inertia of the star. The superfluid Ekman number is approximately
\begin{eqnarray}
 E_s &=&3.06\times 10^{-17} \left( \frac{\nu_s}{3.06\times 10^{-3}\, {\rm cm^2\,s^{-1}}} \right) \left( \frac{\Omega}{100\,{\rm rad\,s^{-1}}} \right)^{-1} \left(\frac{L}{10^6\,{\rm cm}}\right)^{-2}\,, \label{eq5.9}
\end{eqnarray}
Under these assumptions (\ref{eq5.7}) can be written
\begin{equation}
 f(t)=e^{-t/t_d}\left[\cos\left(\frac{2 \pi t}{t_p}\right)+\frac{t_p}{2 \pi t_d}\sin\left(\frac{2 \pi t}{t_p}\right)\right]\,, \label{eq5.12}
\end{equation}
where
\begin{eqnarray} 
  t_d&=&\frac{1}{\Omega |{\rm Re}(s_\pm)|}\approx \frac{e^{\beta A}}{\Omega \beta_0} \,, \label{eq5.10} \\
  t_p&=&\frac{2\pi }{\Omega |{\rm Im} (s_\pm)| } \approx\frac{\pi e^{\beta A/2}}{\Omega}  \sqrt{\frac{\rho k}{K \rho_s}} \,. \label{eq5.11}
\end{eqnarray}

\begin{figure*}
  \includegraphics[width=0.5\textwidth]{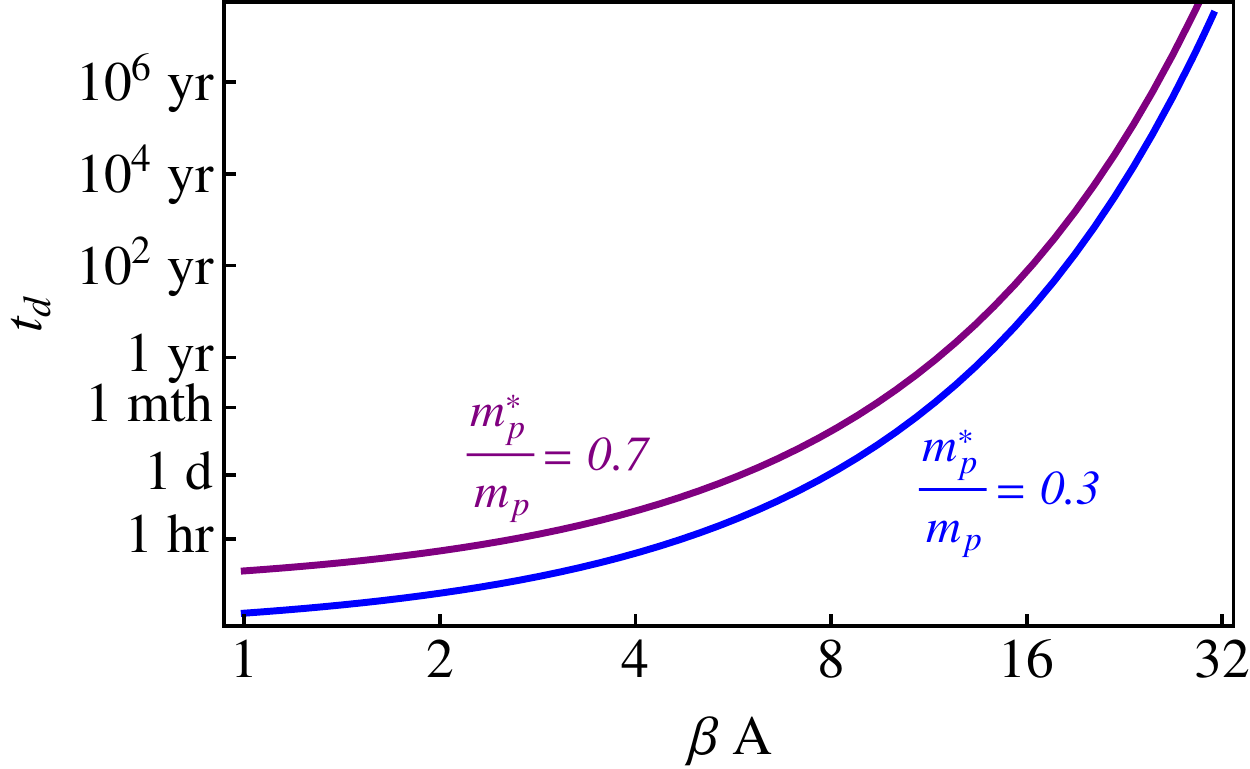}
  \includegraphics[width=0.5\textwidth]{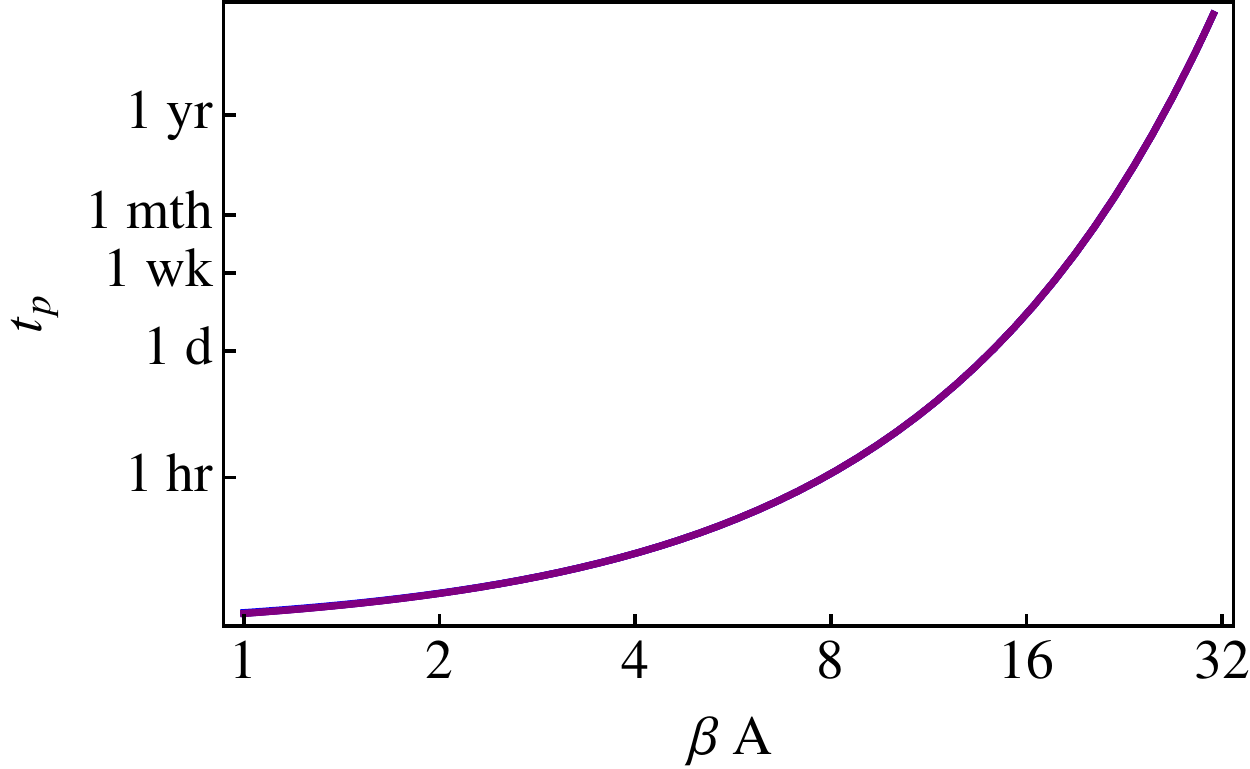}
\caption{Damping time (left) and period (right) for oscillations in a neutron star.  Lines are plotted for $y = m_p^*/m_p=0.3$ (blue) and $0.7$ (purple).  In the right panel the period is independent of $m_p^*$ and the lines overlap.  }
\label{fig6}    
\end{figure*}

In Figure \ref{fig6} we plot the damping time and period for superfluid oscillations in a neutron star as a function of the pinning parameter $\beta A$.  We plot the full result for the damping time and period using (\ref{eq5.8}).
Results are plotted for $y = m_p^*/m_p=0.3$ (blue) and $0.7$ (purple).  
We find that the approximate results given in (\ref{eq5.10}) and (\ref{eq5.11}) give excellent agreement.  The damping time becomes longer as pinning is increased, because the amount of vortices moving and dissipating energy through scattering is exponentially suppressed.  The period also increases as pinning increases, because pinning is inhibiting the movement of vortices and hence the transport of angular momentum.  In the limit of infinitely strong pinning, the vortices stop moving and the period becomes infinitely long.

In principle, oscillations of almost any period are possible, depending on the pinning strength, $A$. The oscillation amplitude is unaffected by pinning and always equal to the glitch amplitude.
The damping time-scale is always longer than the period, so the oscillations are only weakly damped and the second term in (\ref{eq5.12}) is small.  
Post-glitch oscillations are observed in both the Crab and Vela pulsars that are candidates for the oscillatory Ekman pumping mechanism in this paper.  In the Crab, oscillations with a period of roughly four months were reported following the earliest glitches \citep{rud70a,rud70b}.  The ``overshoot'' observed in the recovery the 1975 and 1986 Crab glitches could also correspond to a damped oscillation with period of 200-300 days \citep{won01,van10}. The former would require $\beta A \approx 23.8$, and for the latter $\beta A\approx 24.8 $. 
In Vela, a damped periodic oscillation is clearly visible in the timing residuals following the 1988 glitch [see Figure 2 of \citet{mcc90}]. These oscillations appear to have a period of $\sim 20\, {\rm days}$, requiring $\beta A\approx 20.2 $.
Therefore, superfluid Ekman pumping can explain these oscillations if pinning between vortices and flux tubes is extremely strong, limiting vortices to a very slow creep.

These numbers can be compared with theoretical expectations using known parameters in the outer core of the pulsar.  Assuming a typical pulsar temperature and using parameter estimates from \citet{lin14a}, from (\ref{eq5.5a}) we find
\begin{equation}
  \beta A= 2.05 \times 10^5 \left(\frac{100\,{\rm MeV}}{E_p}\right)^{1/2}
\left(\frac{0.12}{\mathcal{T}}\right)^{1/2}\left(\frac{10^7\,{\rm K}}{T}\right)^{-1}\,, \label{eq5.16}
\end{equation}
where we have taken $\Delta \omega =0$.  Clearly (\ref{eq5.16}) leads to extremely long oscillation periods and damping times.  However, this is a gross upper estimate because the lag $\Delta \omega$ is not likely to be zero.  If the lag is close to the critical lag for unpinning $\Delta \omega_c$, the estimate (\ref{eq5.16}) can be greatly reduced, see (\ref{eq5.5}).  However, to correctly determine the lag the non-linear calculation including velocity dependence in the mutual friction coefficients is required, which is beyond the scope of this paper.

Finally, we remark on limitations of the present model.  Neutron stars are strongly stratified, typically with buoyancy frequencies $N$ exceeding the rotation rate \citep{rei92}.  Buoyancy inhibits Ekman pumping from penetrating the core, confining the circulation to a buoyancy layer with thickness $\Omega L/N $ \citep{wal69,abn96,van08a}.  The spin-up is time is reduced commensurately reduced to $E^{-1/2} N^{-1}$ because the effective size of the Ekman cell has been reduced by a fraction $\Omega /N$.  The final state is not co-rotation, but has persistent shear between the fluid in the buoyancy layer and that in the core which in untouched by Ekman pumping \citep{mel12}.  In a typical neutron star, we find $\Omega/N \sim 0.20$ \citep{rei92}, so that the spin-up time is reduced by a factor of 5.

A similar result is expected for the superfluid Ekman pumping mechanism discussed in this paper.  As in the classical stratified Ekman pumping, we expect that buoyancy forces will confine the circulation to a depth $\Omega L/N $. The Taylor-Proudman theorem no longer applies in the core, and the vortex lines will be sheared.  Above $\Omega L/N $ the fluid undergoes in superfluid oscillations, but at greater depths the fluid does not participate.  To estimate the period of oscillations, we assume that the length scale for the Ekman cell is reduced from $L$ to $\Omega L/N $, and (\ref{eq3.26}) becomes
\begin{equation}
  t_P=\frac{\pi}{\Omega}\sqrt{\frac{\Omega}{N}\left( \frac{k}{1+K} \right) }\,.
\end{equation}  
In this case we have $\sqrt{\Omega/N}\sim 0.45$, shortening the period by a factor of approximately of 2.2.  Therefore the superfluid oscillations are more weakly affected by stratification than classical Ekman pumping.

\section{Conclusions} \label{sec6}

The response of a uniformly rotating superfluid threaded with a large density of vortex lines to an impulsive response of its container has been investigated.  When vortices are strongly pinned to the container and there is no dissipation, the system oscillates persistently with a period of order $E^{-1/4}\Omega^{-1}$, where $E_s$ is the dimensionless vortex line tension parameter.  
The low-frequency mode is generated by a secondary flow analogous to classical Ekman pumping, periodically reversed by the vortex tension.  
This secondary flow transports vortices radially inwards and outwards in the interior, changing the angular momentum of the fluid.
The full set of equations of \citet{cha86} have been solved and compared with experiments performed by \citet{tsa80} in superfluid helium.  We find qualitative agreement between theory and experiment, however quantitative comparison is difficult because some experimental parameters are unknown, and the experiment may probe the non-linear regime, which is beyond the scope of this theory.  The columnar nature of superfluid Ekman pumping also reproduces the experimentally reported dependence of the frequency on vessel radius and height, however the theory would benefit from comparison with new, carefully conducted experiments.  We argue that Tkachenko modes are unlikely to be excited by the boundary and initial conditions imposed in the spin up experiments of \citet{tsa80}.  
In neutron stars, the movement of neutron vortices is inhibited by pinning to type II superconducting protons in the core, slowing the superfluid Ekman pumping process.  For very strong pinning, and hence slow vortex creep, the theory can explain post-glitch oscillations with periods of days to years, with much longer damping times.   \newline

The author thanks Bennett Link for useful feedback on the manuscript.

\appendix
\section{General solution in the boundary layer approximation}\label{secA1}

Here we present the solution to the governing equations (\ref{eq4.2})--(\ref{eq4.9}).  The solution is too large to present algebraically, so we outline the method and present key results.  Because if the size of the equations, it is easier to perform the calculations in programs such as Mathematica or Maple and handle them numerically.

We assume the container is rigid so that the upper and lower boundaries are move identically.  Therefore, the $V_{n,s}$ are symmetric in $z$ and $\chi_{n,s}$ are anti-symmetric.  Applying these boundary conditions, we solve in the upper-half plane only.

The normal fluid has the solution to the interior flow equations (\ref{eq4.2})--(\ref{eq4.5})
\begin{eqnarray}
  V^I_{n}(t)=\sum_{i=i}^3 V^I_{ni}e^{s_i t} +V^I_{n4}\,, \label{eqA1}
\end{eqnarray}
and similarly for $V_s$.  Three time-scales exist in the problem and are obtained when applying the boundary conditions below.  The stream-function for the secondary flow is
\begin{eqnarray}
  \chi^I_{n}(z,t)=z\left[\sum_{i=i}^3 \chi^I_{ni}e^{s_i t} +\chi^I_{n4} \right]\,, \label{eqA2}
\end{eqnarray}
and similarly for $\chi_s$.  
The pressure is
\begin{eqnarray}
  P_{n}(t)=\sum_{i=i}^3 P_{ni}e^{s_i t} +P_{n4}\,, \label{eqA3}
\end{eqnarray}
and similarly for $P_s$.  
Substituting (\ref{eqA1})--(\ref{eqA3}) into (\ref{eq4.2})--(\ref{eq4.5}) allows to solve for the coefficients of $\chi_{n,s}$ and $P_{n,s}$ in terms of those of $V^I_{n,s}$.

In the boundary layer, the solutions to (\ref{eq4.6})--(\ref{eq4.9}) have the form
\begin{eqnarray}
  V^B_n(z,t)&=&\sum_{i=1}^3e^{s_i t} \left[V^B_{ni+}e^{-k_+(1-z)}+V^B_{ni-}e^{-k_-(1-z)} \right]\,, \label{eqA4} \\
 V^B_s(z,t)&=&\sum_{i=1}^3e^{s_i t} \left[V^B_{si+}e^{-k_+(1-z)}+V^B_{si-}e^{-k_-(1-z)}+V^B_{si}e^{-k(1-z)} \right]\,,  \label{eqA5}
\end{eqnarray}
in the upper boundary layer, where 
\begin{eqnarray}
  k_\pm&=&\sqrt{\frac{\pm 2 i \rho \left[B\pm i \left(2 -B'\right)\right]}{E\left[\rho_n B\pm i \left(2  \rho - \rho_n B' \right)\right]}}\,,   \label{eqA6} \\
 k&=&\sqrt{\frac{2}{E_s}}\,.  \label{eqA7}
\end{eqnarray}
For $\chi^B_{n,s}$ we have
\begin{eqnarray}
  \chi^B_n(z,t)&=&\sum_{i=1}^3e^{s_i t} \left[\chi^B_{ni+}e^{-k_+(1-z)}+\chi^B_{ni-}e^{-k_-(1-z)} \right]\,, \label{eqA8}  \\
 \chi^B_s(z,t)&=&\sum_{i=1}^3e^{s_i t} \left[\chi^B_{si+}e^{-k_+(1-z)}+\chi^B_{si-}e^{-k_-(1-z)}+\chi^B_{si}e^{-k(1-z)} \right]\,.  \label{eqA9}
\end{eqnarray}
Note that for the classical Ekman boundary layers (with exponents $k_\pm$), the superfluid and normal fluid components are coupled, whereas, for the superfluid Ekman pumping mechanism, the normal fluid is not involved.
Substituting (\ref{eqA4})--(\ref{eqA9}) into (\ref{eq4.6})--(\ref{eq4.9}) and comparing like terms gives $V^B_{si\pm}$, $\chi^B_{si\pm}$ and $\chi^B_{ni\pm}$ in terms of $V^B_{ni\pm}$.  The coefficients $V^B_{si}$ and $\chi_{si}^B$ are unrelated to other boundary layer coefficients.

The motion of the container has the solution
\begin{eqnarray}
  f(t)=\sum_{i=i}^3 f_{i}e^{s_i t} +f_{4}\,, \label{eqA10}
\end{eqnarray}
where the $f_i$ can be determined in terms of $V^B_{ni\pm}$, $V^B_{si}$ and $\chi{si}^B$ by substituting (\ref{eqA4})--(\ref{eqA9}) into (\ref{eq2.32}).  Note that only the boundary layer corrections to the velocity contribute to the torque on the container.

The coefficients $V^B_{ni\pm}$, $V^B_{si}$ and $\chi_{si}^B$ and $V^I_{si}$ and $V^I_{s4}$ can be solved for in terms of $V^I_{ni}$ through application of the boundary conditions (\ref{eq2.37}), (\ref{eq3.41}) and (\ref{eq2.39}) for the superfluid and (\ref{eq4.10})--(\ref{eq4.12}) for the normal fluid and comparing like terms.  For each $i$, the six boundary conditions determine five unknowns in terms of $V^I_{ni}$, the final equation is an eigenvalue equation for $s_i$.  The equation is cubic is $s_i$ giving solutions for the three $s_i$.
The boundary conditions also give $f_4=V_{s4}=V_{n4}$.

The remaining four unknown coefficients $V^I_{ni}$ and $V_{n4}$ are determined by the initial conditions (\ref{eq3.42})--(\ref{eq3.44}) and (\ref{eq4.13}).  The result is too cumbersome to present algebraically, but is presented for the experimentally measured parameters in helium II in equation (\ref{eq4.14}).

\section{Solutions in the boundary layer approximation for strong coupling}\label{secA2}

In helium II, we have $B,B'\sim 1$, resulting in the strong coupling between the two fluid components.  Any differential rotation is removed over the rotational time-scale and the two fluid-components are ``locked together'' over the much longer Ekman time \citep{rei93,van13}.  This assumption can be used to find a more analytically tractable solution to that presented in \S\ref{secA1}.

When $B=B'=0$, (\ref{eq4.3}) and (\ref{eq4.5}) describe the geostrophic balance in the interior, balancing the time rate of change of the azimuthal flow with secondary flow. In (\ref{eq4.3}) all terms are of order $E^{1/2}$ \citep{gre63} and in (\ref{eq4.5}) all terms are order $E_s^{1/4}$ (see \S\ref{sec2}).  However, when $B, B'\sim 1$ the third terms in (\ref{eq4.3}) and (\ref{eq4.5}) are order unity, while the remainder are of the order of the secondary flow.  Therefore, we perturb the leading order velocity so that to leading order we have $V^I_n=V^I_s$ and the velocity difference $\delta V^I$ is of the order of the secondary flow, i.e., 
\begin{eqnarray}
  V^I_s=V^I_n-\delta V^I\,.
\end{eqnarray}
Equations (\ref{eq4.2})--(\ref{eq4.5}) become
\begin{eqnarray}
0&=& 2 V^{I}_n - P_n \, , \label{eqB2} \\
0&=&\frac{\partial V^{I}_n}{\partial t}+ 2\frac{\partial \chi^I_n}{\partial z} + \frac{\rho_s B}{\rho}\delta V^{I}-\frac{\rho_s B'}{\rho} \left( \frac{\partial \chi^I_n}{\partial z} - \frac{\partial \chi^I_s}{\partial z} \right) \, ,  \label{eqB3} 
\end{eqnarray}
\begin{eqnarray}
0&=&2 V^{I}_s  - P_s \, , \label{eqB4}\\
0&=& \frac{\partial V^{I}_s}{\partial t} +2  \frac{\partial \chi^I_s}{\partial z}-\frac{\rho_n B}{\rho}  \delta V^{I} + \frac{\rho_n}{\rho} B'  \left( \frac{\partial \chi^I_n}{\partial z}- \frac{\partial \chi^I_s}{\partial z}\right) \, , \label{eqB5} 
\end{eqnarray}
The solution for $V^I_n$ has the same form (\ref{eqA1}), and similarly for $\delta V^I$ we have
\begin{eqnarray}
  \delta V^I (t)=\sum_{i=i}^3 \delta V^I_{i}e^{s_i t} + \delta V^I_{4}\,, \label{eqB6}
\end{eqnarray}
Using the same form for the stream-functions (\ref{eqA2}) and pressures (\ref{eqA3}), their coefficients can be determined in terms of $V_I^{ni}$ and $\delta V^I_i$ using (\ref{eqB2})--(\ref{eqB5}).

The remaining procedure follows as in \S\ref{secA1}, however we find that there are now only two time-scales $s_1$ and $s_2$; the third corresponded to mutual friction coupling which is now effectively instantaneous. Accordingly, we can no longer impose separate initial conditions on $V_s^I$ and $V^I_n$, which must be the same.
The resulting solutions are still algebraically cumbersome, so we present results in the limit of negligible viscosity and negligible vortex tension.

\subsection{Solution for negligible viscosity} \label{secA2.1}

A useful result discussed in \S\ref{sec4.2} is the case when viscous effects are negligible.  We argue in that section that this is relevant in Helium II.

In the boundary layer we find that $V^B_{n,si\pm}=0$ and $\chi^B_{n,si\pm}=0$.  We also do not need to satisfy the normal fluid boundary conditions for no slip (\ref{eq4.10}) and (\ref{eq4.12}).  The result is
\begin{eqnarray}
  f(t)&=& \frac{1}{1+K}\left[1+ K \frac{s_1 e^{s_2 t}-s_2 e^{s_1 t}}{s_1-s_2}\right]\,, \\
 V^I_n&=& \frac{1}{1+K}\left[1- \frac{s_1 e^{s_2 t}-s_2 e^{s_1 t}}{s_1-s_2}\right]\,, \\
 \delta V^I&=&\frac{2\left(e^{s_2 t}-e^{s_1 t}\right)}{\beta k \left(s_1-s_2\right)}\,, \\
 \chi^I_s&=&\frac{2 z \left(e^{s_2 t}-e^{s_1 t}\right)}{k \left(s_1-s_2\right)}\,, \\
 \chi^I_n&=&0\,,
\end{eqnarray}
where
\begin{equation}
  s_{1,2}=-\frac{ 1}{k \gamma}\left[ 1\pm \sqrt{1-\frac{4\left(1+K\right)\gamma^2 k \rho_s}{ \rho }} \right]\,. \label{eq4.16}
\end{equation}
and
\begin{equation}
  \beta=\frac{B}{2-B'}\,.
\end{equation}
The implications of this result are discussed in \S\ref{sec4.2}

\subsection{Solution for negligible vortex pinning or vortex tension} \label{secA2.2}

The focus of this paper has been on pinned vortices, however, as a useful check we can recover the previous results of \citet{rei93} and \citet{van13,van14a} where pinning is negligible.  

When $\gamma\rightarrow 0$, we obtain the result
\begin{eqnarray}
  f(t)&=& \frac{1}{1+K}\left(1+ K  e^{s_3 t} \right)\,, \\
 V^I_n&=& \frac{1}{1+K}\left(1- e^{s_3 t}\right)\,, \\
 \delta V^I&=&\frac{\rho e^{s_3 t}\left(k_1^2+k_2^2\right)\left[\left(k_1+k_2\right)+ i \beta \left(k_1-k_2\right)\right]}{4 \beta^2 k_1^2 k_2^2 \rho_s}\,, \\
 \chi^I_n&=&\frac{i z e^{s_3 t}\left(k_1-k_2\right)}{2 k_1 k_2}\,, \\
 \chi^I_s&=&\chi^I_n-\frac{\rho e^{s_3 t}\left(k_1^2+k_2^2\right)\left[\left(k_1+k_2\right)+ i \beta \left(k_1-k_2\right)\right]}{4 \beta^2 k_1^2 k_2^2 \rho_s}\,,
\end{eqnarray}
where
\begin{equation}
  s_3=-\frac{E}{2} \left(k_1+k_2\right)\left(1+K\right)\rho_n \,. \label{eqB19}
\end{equation}
Interestingly the same result is obtained by taking $E_s=0$.  This is just a statement that when there is no vortex pinning, the vortex tension plays no active role in the spin-up, as discussed in \citet{van13}.  There is a slight difference in the boundary layer solution; for $E_s=0$, $V^B_{s3}=\chi^B_{s3}=0$, while for $\gamma=0$ they are non-zero.

From (\ref{eqB19}) we obtain the spin-up time
\begin{eqnarray}
 t_{\rm s} &=&\frac{1}{\Omega |s_3|}=\frac{\rho}{\rho_n  E^{1/2}\Omega \left(1+K\right) } \nonumber \\
 &\times& \left\{ \rho \left[\frac{B^2+\left(2-B'\right)^2}{\left(\rho_n B\right)^2+\left( 2 \rho-\rho_n B' \right)^2}\right]^{1/2} + \frac{2 \rho_s \rho B}{\left(\rho_n B\right)^2+\left(2\rho-\rho_n B'\right)^2}\right\}^{-1/2}\,. \label{eq83}
\end{eqnarray}
This result was obtained by \citet{rei93} and explored in detail by \cite{van13}.  The term in the curly braces arises from the coupling between the two fluid components is a boundary layer of order unity.  The pre-factor is the familiar Ekman time modified by the normal fluid density fraction.  The two fluid components are strongly coupled (i.e., they achieve co-rotation within a rotation period as $B\sim 1$), but it is only the normal fluid providing the Ekman pumping, which is reflected in the spin-up time.  

\bibliographystyle{jfm}

\bibliography{references}

\end{document}